\documentclass[%
 reprint,
 amsmath,amssymb,
 prapplied,
floatfix,
]{revtex4-2}
\usepackage{todonotes}
\usepackage{comment}
\usepackage{subcaption}
\usepackage{algorithm}
\usepackage{caption}
\usepackage{algpseudocode}
\usepackage{booktabs}
\usepackage{derivative}
\usepackage{graphicx}% Include figure files
\usepackage{dcolumn}% Align table columns on decimal point
\usepackage{bm}% bold math
\usepackage{amsfonts,amsmath,amssymb,amsthm}
\usepackage{siunitx}
\usepackage{pifont}
\usepackage{multirow, mathtools}

\newtheorem{proposition}{Proposition}
\newtheorem{corollary}{Corollary}
\theoremstyle{definition}
\newtheorem{definition}{Definition}
\newtheorem{example}{Example}
\newcommand{\norm}[1]{\left\lVert #1\right\rVert}

\newcommand{\onenorm}[1]{\left\lVert #1\right\rVert_1}
\newcommand{\kl}[2]{\mathrm{D}_{\mathrm{KL}}(#1\Vert #2)}
\newcommand{\innerprod}[2]{\left\langle#1, #2\right\rangle}
\newcommand{\R}{\mathbb{R}}
\newcommand{\E}{\mathbb{E}}

\newcommand{\excl}[1]{{\backslash \hspace{-0.3em} #1}}
\begin{document}

\preprint{APS/123-QED}

\title{Limitations in Parallel Ising Machine Networks: Theory and Practice}% Force line breaks with \\

\author{Matthew X. Burns}
 \email{mburns13@ur.rochester.edu}
\author{Michael C. Huang}%
 \email{michael.huang@rochester.edu}
\affiliation{%
 Department of Electrical and Computer Engineering\\
 University of Rochester
}%

\date{\today}% It is always \today, today,
\newcommand{\hamil}{\mathcal{H}}
\begin{abstract}
Analog Ising machines (IMs) occupy an increasingly prominent area of computer architecture research, offering high-quality and low latency/energy solutions to intractable computing tasks. However, IMs have a fixed capacity, with little to no utility in out-of-capacity problems. Previous works have proposed parallel, multi-IM architectures to circumvent this limitation~\cite{sharma_increasing_2022,santos_enhancing_2024}. In this work we theoretically and numerically investigate tradeoffs in parallel IM networks to guide researchers in this burgeoning field. We propose formal models of parallel IM excution models, then provide theoretical guarantees for probabilistic convergence. Numerical experiments illustrate our findings and provide empirical insight into high and low synchronization frequency regimes. We also provide practical heuristics for parameter/model selection, informed by our theoretical and numerical findings.
\end{abstract}

\begin{comment}
    General Flow of Article:
    - Intro and BG on IMs/Analog Optimizers
    - (Maybe) Mathematical Preliminaries
    - Description of Communication Model
    - 
\end{comment}
% General Flow of Article

% Intro
% BG on Ising Machines/Analog Optimizers
% Description of Communication Model
%   

\maketitle
\section{Introduction}\label{sec:intro}

With the slowing of Moore's law, dynamics-based analog computation offers an alternative path to increased performance and power efficiency. Recent dynamical accelerator proposals include adiabatic quantum computing (AQC)~\cite{albash_adiabatic_2018}, quantum annealing (QA)~\cite{hauke_perspectives_2020}, analog electronics~\cite{afoakwa_brim_2021,wang_oim_2019} and photonic~\cite{inagaki_coherent_2016} systems: each offering low-latency computation for specific tasks. Furthermore, classical analog optimization (electronics and photonics) are order-of-magnitude more energy efficient than traditional systems, prompting further interest to lower the energy footprint of high-performance computing. The most prominent class relies on continuous relaxations of the Ising spin glass Hamiltonian, earning the appellation ``Ising Machines'' (IMs)~\cite{mohseni_ising_2022,zhang_review_2024}. Simulated behavior and hardware prototypes have demonstrated that Ising machines can effectively accelerate a number of NP-Hard combinatorial tasks, including graph maximum cut~\cite{afoakwa_brim_2021,inagaki_coherent_2016,wang_oim_2019}, LDPC decoding~\cite{elmitwalli_utilizing_2024}, satisfiability testing~\cite{sharma_augmenting_2023,tan_hyqsat_2023}, energy-based model inference~\cite{vengalam_supporting_2023}, and data clustering~\cite{matsumoto_distance-based_2022-2}.

However, current Ising machine proposals have a number of limitations: chief among which is a fixed device capacity. Once designed, an IM has a fixed number of degrees of freedom. If that limit is exceeded, the IM alone cannot solve the problem directly. Broadly speaking, there are two approaches to the problem: \ding{172} use the IM hardware as a subsolver in an otherwise conventional algorithmic framework; or \ding{173} orchestrate multiple IM chips to form a logically larger IM. 

\textbf{The subsolver approach:} Conceptually, the overall problem is still solved conventionally by a digital algorithm. Except IM hardware is available to accelerate subproblem solving. Note that a naive divide-and-conquer technique generally fail~\footnote{When solving spin glass Hamiltonians (and the combinatorial optimization problems they define) using naive divide-and-conquer, the need for digital pre- and post-processing for each subproblem quickly dominates the workload, limiting the potential speedup and incurring high energy costs from the added communication and processing.}. The current state of the art is the D-Wave Leap Hybrid framework, in which quantum annealing ``guides [a] classical heuristic''~\cite{mcgeoch_d-wave_2020}, such as Tabu search or simulated annealing (SA). While some works have shown time-to-solution advantages in crafted benchmark problems~\cite{mcgeoch_d-wave_2020}, hybrid solvers can still be outperformed by classical algorithms~\cite{du_new_2025}. Works targeting problem-specific applications such as network community detection~\cite{kalehbasti_ising-based_2021} and satisfiability~\cite{tan_hyqsat_2023} have achieved some limited speedup, however their approach is both highly limited by Amdahl's law and are (by design) not general-purpose. 

\begin{figure}[h]
    \centering
    \includegraphics[width=\linewidth]{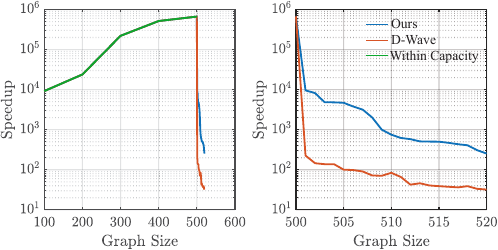}
\caption{From Ref.~\cite{sharma_increasing_2022}. Estimated speedup degradation as solver capacity (500 spins) is exceeded. The right plot is a zoomed-in version of the left. ``Ours'' refers to an annealing-based divide-and-conquer scheme tested in Ref.~\cite{sharma_increasing_2022}, while ``D-Wave'' refers to the \texttt{qbSolv}~\cite{booth_partitioning_nodate} framework.}
    \label{fig:anshujit_plot}
\end{figure}
\begin{figure}[h]
    \centering
    \includegraphics[width=\linewidth]{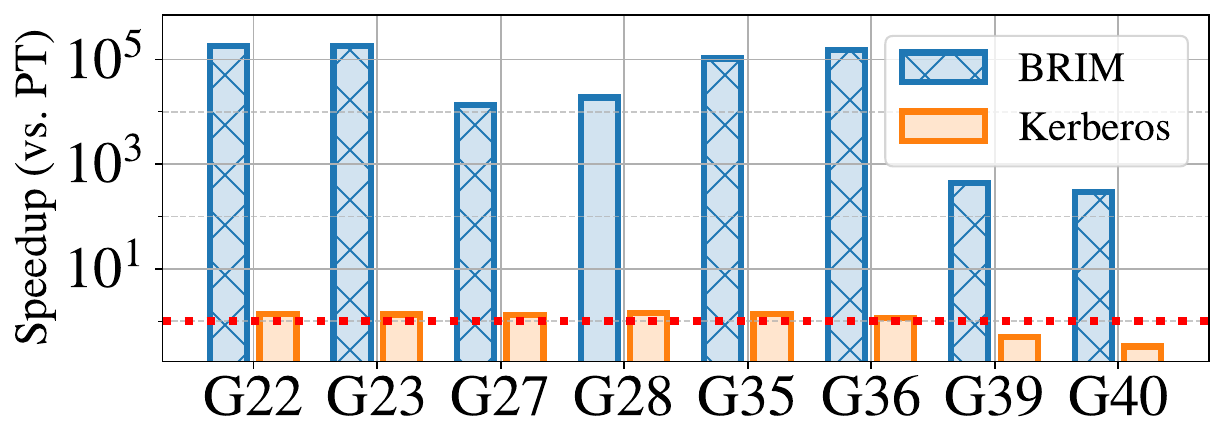}
\caption{Estimated IM speedup versus CPU-based parallel tempering (PT) running on 8 cores on problems from the GSet suite~\cite{noauthor_index_nodate}. Simulated BRIM~\cite{afoakwa_brim_2021} behavior represents a typical analog Ising machine, while D-Wave Kerberos~\cite{reference_workflows_dwave} represents a typical hybrid platform. All solvers are run to a minimum target of 99\% of the best-known-solution.}
    \label{fig:dwave_bad}
\end{figure}

Experimental results have demonstrated the limitations of the subsolver approach. Ref~\cite{sharma_increasing_2022} examined the scaling behavior of ``divide-and-conquer'' hybrid solvers. Fig.~\ref{fig:anshujit_plot} (from~\cite{sharma_increasing_2022}) shows the estimated speedup over CPU-based simulated annealing using the (then state-of-the-art) \texttt{qbSolv}~\cite{booth_partitioning_nodate} framework (``D-Wave'') and a simulated annealing based solver examined in Ref.~\cite{sharma_increasing_2022} (``Ours''). The speedup is shown to decay precipitously as the problem size barely exceeds hardware capacity, motivating the authors to pursue hardware-only approaches.
Fig.~\ref{fig:dwave_bad} reinforces the conclusions of Ref~\cite{sharma_increasing_2022} using the more modern D-Wave Kerberos hybrid workflow~\cite{reference_workflows_dwave}. Here we \ding{172} simulate a typical analog IM (BRIM~\cite{afoakwa_brim_2021}) and \ding{173} run Kerberos using the D-Wave Leap Hybrid service, with problems selected from the 2000-node GSet suite graphs~\cite{noauthor_index_nodate}. Both solver times are then normalized to CPU-based parallel tempering~\cite{jearl_parallel_2005}. We again see orders of magnitude differences in performance, highlighting the significant challenge of the subsolver approach to IM scalability.

\textbf{The multi-chip approach} attempts to preserve energy efficiency and solver speed at the cost of extra IM hardware. The problem is partitioned into disjoint subproblems, each solved on a different IM communicating over a common network. So-called ``multi-chip'' approaches are a mainstay of traditional computer architecture, but run into significant difficulties when applied to analog systems. As shown empirically by Sharma et al. 2022, communication bandwidth is the limiting factor in both speedup and solution quality. The authors then explored two solution paths: \textit{serial} and \textit{concurrent} execution models. Serial execution adopts a pipelined execution model which sacrifices latency for bandwidth savings, while concurrent execution provides lower latency operation with higher communication costs. The tradeoffs were empirically studied using Sherrington-Kirkpatrick spin glasses, but not discussed in generality. In particular, the behavior and convergence properties of concurrently executing subsystems are not well understood within the IM community.

% P 4: In this work...
In this work, we perform an in-depth theoretical and numerical study on parallel Ising machine architectures, with a particular emphasis on lesser studied concurrent execution models. We focus here on Ising machines whose behavior can be modeled by an overdamped Langevin equation. Our contributions can be summarized as
\begin{enumerate}
    \item Convergence analysis for concurrent execution mode, providing lower bounds in KL-divergence (Proposition~\ref{prop:kl}) and sufficient conditions for per-iteration contraction (Propositions~\ref{prop:w1_1} and~\ref{prop:w1_2});
    \item Numerical results explicitly demonstrating our theoretical findings on small, enumerable systems;
    \item Energy and time efficiency comparisons between parallel execution models across a range of network bandwidths and energy costs in canonical MaxCut benchmarks;
    \item Practical, application-dependent heuristics for execution model and parameter selection.
\end{enumerate}

Section~\ref{sec:prelims} provides an overview of relevant topics used in our analysis. Section~\ref{sec:model} proposes a general framework for parallel Ising machine analysis: applicable to a wide swath of current proposals. Section~\ref{sec:convergence} provides theoretical convergence properties. Narrowing our focus to linear Ising machines~\cite{afoakwa_brim_2021}, Section~\ref{sec:experiments} provides numerical experiments comparing the time and energy costs of each proposal in unconstrained binary optimization, then proposes a decision tree for choosing between serial and parallel execution.
Section~\ref{sec:conclusion} concludes by summarizing our findings.

\section{Preliminaries}\label{sec:prelims}
\subsection{Combinatorial Optimization by Spin Glass Sampling}
\newcommand{\spinvec}{\bm{s}}
\newcommand{\spin}{s}
We begin by recalling the general Ising spin glass Hamiltonian $\hamil(\spinvec)$:
\begin{equation}
    \hamil(\spinvec)=-\sum_{i\neq j}J_{ij}\spin_i\spin_j-\sum_{i=1}^N h_i\spin_i
\end{equation}
where $\spinvec\in\{-1,1\}^N$ is a vector of 1D spins, $J\in \mathcal{S}(\mathbb{R})^{N\times N}$ is an $N\times N$ symmetric matrix describing spin-spin couplings, and $h\in \mathbb{R}^N$ is an external field term. 

Spin glass systems are typically studied within the canonical ensemble with equilibrium statistics described by the Gibbs-Boltzmann distribution $\pi(\spinvec)$:
\begin{equation}
    \pi(\spinvec)=\frac{1}{Z}\exp[-\beta\hamil(\spinvec)]
\end{equation}
where $\beta=T^{-1}$ is the inverse temperature of the heat bath with the Boltzmann constant set to unity for convenience. In the limit $\beta\to\infty$ $\pi(\spinvec)$ concentrates around the ground states of $\hamil(\spinvec)$ 

Finding the ground state of a spin glass system with non-planar coupling graphs\footnote{The coupling matrix $J$ equivalently describes a weighted graph with spins as nodes.} is NP-Hard~\cite{barahona_computational_1982,istrailStatisticalMechanicsThreedimensionality2000}, meaning that efficient spin glass optimizers can be used to solve any problem in the NP complexity class. Moreover, several real-world NP-hard optimization problems have equivalent Ising formulations~\cite{lucas_ising_2014}, including integer linear programming, production scheduling, and graph coloring (to name a few). 

The classic unconstrained COP is the (weighted) maximum cut (MaxCut) problem. Given a graph $G=(V,E)$ with weighted adjacency matrix $W$, the MaxCut problem seeks to partition the graph into two sets $S$, $T$ such that the weight of edges crossing the partition is maximized. MaxCut has a trivial representation in as an Ising Hamiltonian with $J=-W$ with spin values denoting set membership. Given the energy $\hamil(\spinvec)=-\frac{1}{2}\spinvec^TJ\spinvec$, the cut value $C(S,T)$ can be computed as
\begin{equation}
    C(S,T)=\frac{1}{2}\left(\sum_{(u,v)\in E}W_{uv}-\hamil(\spinvec)\right).
\end{equation}

MaxCut is a widely used benchmark task for Ising machines, however real-world problems typically have constraints defining feasible solutions. Constrained optimization is generally accomplished by introducing a penalty term $\hamil_{Con}(\spinvec)$ with multiplier $\gamma$
\begin{equation}
    \hamil(\spinvec)=\lambda \hamil_{Con}(\spinvec)+\hamil_{Obj}(\spinvec).
\end{equation}
$\hamil_{Obj}$ represents the optimization objective, free of any constraints, while $\hamil_{Con}(\spinvec)$ penalizes infeasible states. A canonical constrained combinatorial problem is the Traveling Salesman Problem (TSP). Given $N$ cities with $N\times N$ distance matrix $D$, the TSP seeks a Hamiltonian path~\footnote{A route such that each city is visited exactly once} of cities which minimizes the total distance traveled. $\lambda$ must be chosen sufficiently large to ensure that it is generally unfavorable to violate problem constraints to improve the objective function. For the best performance, it is typically tuned per problem, however some problem-agnostic heuristics have successfully recovered high-quality solutions~\cite{ayodele_penalty_2022}.

\subsection{Comparing Probability Measures}\label{sec:prelims_dist}
Instead of directly finding the ground state, we can alternatively attempt to define a random process whose sample distribution converges to $\pi$. Accordingly, we need to define notions of ``distance'' or ``divergence'' between two distributions. Note that throughout this work we abuse notation by identifying a probability distribution with its density w.r.t. the Lebesgue measure.

Comparisons between two probability measures $\mu$ and $\nu$ often arises when observing stochastic processes, hence a number of functionals exist which quantify proximity in distribution. In this work, we make use of two. The first is the $p$-Wasserstein distances $W_p$,
\begin{equation}
    \label{eqn:wpdef}W_p(\mu,\nu)=\inf_{\gamma\in\mathcal{C}(\mu,\nu)}(\mathbb{E}_\gamma||x-y||_p^p)^{1/p}
\end{equation}
where $\gamma$ is a coupling~\footnote{A probability distribution over the joint space $X\times Y$ satisfying $\int_X\gamma dx=\nu$ and $\int_Y\gamma dy=\mu$} over the measures $\mu$ and $\nu$ and $||\cdot||_p$ is the $p$-norm. The Wasserstein distance measures the minimal amount of ``work'' needed to move probability mass from $\mu$ to $\nu$, hence the term ``optimal transport''.  Note that the Wasserstein distance defines a metric over probability distributions, meaning that the measure satisfies a triangle inequality (a property which we will exploit).

Wasserstein distances have seen extensive use in sampling algorithm analysis, particularly for Langevin Monte Carlo methods~\cite{durmus_efficient_2016,dalalyan_user-friendly_2019,li_sharp_2022}. Most works utilize the Euclidean $W_2$, while some analyses focusing on non-convex landscapes have favored $W_1$ due to the relative simplicity of its triangle inequality~\cite{li_sharp_2022}. Recent works in stochastic thermodynamics have made extensive use of optimal transport to bound stochastic entropy production and define optimal control functions~\cite{shiraishi_speed_2018,shiraishi_wasserstein_2024,dechant_minimum_2022,nakazato_geometrical_2021,dechant_thermodynamic_2019}. 

The second measure we will utilize is the Kullbeck-Liebler divergence $\kl{\mu}{\nu}$
\begin{equation}
    \kl{\mu}{\nu}=\mathbb{E}_{\mu}\left[\log\frac{d\mu}{d\nu}\right]
\end{equation}
where $\frac{d\mu}{d\nu}$ is the Radon-Nikodym derivative of $\mu$ w.r.t. $\nu$. Unlike the Wasserstein distance, the $\kl{\mu}{\nu}$ is not a distance, as it is not symmetric and does not satisfy the triangle inequality. Nevertheless, it provides a useful measure of convergence between two distributions, as $\kl{\mu}{\nu}\geq 0$ with equality iff $\mu=\nu$.

In a discrete state space $\Omega$, the expectation takes the form of a sum of the possible system states
\begin{align}
    \label{eqn:wpdisc}&W_p(\mu,\nu)=\inf_{\gamma\in\mathcal{C}(\mu,\nu)}\left(\sum_{x,y\in \Omega}\gamma(x,y)||x-y||_p^p\right)^{1/p}\\
    \label{eqn:kldisc}&\kl{\mu}{\nu}=\sum_{x\in\Omega}\mu(x)\log\frac{d\mu}{d\nu}(x)
\end{align}
\newcommand{\region}{\mathcal{D}}
and integrals in a continuous state space $\region\subseteq \R^d$
\begin{align}
    \label{eqn:wpcont}&W_p(\mu,\nu)=\inf_{\gamma\in\mathcal{C}(\mu,\nu)}\left(\int_{x,y\in \region}||x-y||_p^p\gamma(dx,dy)\right)^{1/p}\\
    \label{eqn:klcont}&\kl{\mu}{\nu}=\int_{\region}\mu(x)\log\frac{d\mu}{d\nu}(x)dx
\end{align}

% \subsection{Stochastic Entropy Production}
% Stochastic thermodynamics 

\section{Problem Partitioning}\label{sec:model}
When the number of spins $N$ exceeds IM capacity $K$, we need to decompose the problem into interacting subsystems, each of which is evolved separately. Without loss of generality we assume that $N$ is an integer multiple of $K$ so that the number of chips needed, $B=N/K$, is an integer, otherwise we can simply pad with non-interacting spins. 

We denote the global problem state $X\in \R^N$, local subsystems as $X_p\in \R^{N_p}$, and individual variables as $x_i\in \R$. Hence $X_p\subseteq X$, $x_{i}\in X_p$. By $\nabla_{X_p}f(X)$, we mean the gradient of function $f$ with respect to system $X_p$, which may have dependencies on other subsystems. By $X_{\setminus p}$, we mean all degrees of freedom \emph{not} in subsystem $X_p$, i.e. $X_{\setminus p}=(X_1,...,X_{p-1}, X_{p+1},...,X_B)$.

\begin{figure*}[th]
    \centering
    \includegraphics[width=0.9\linewidth]{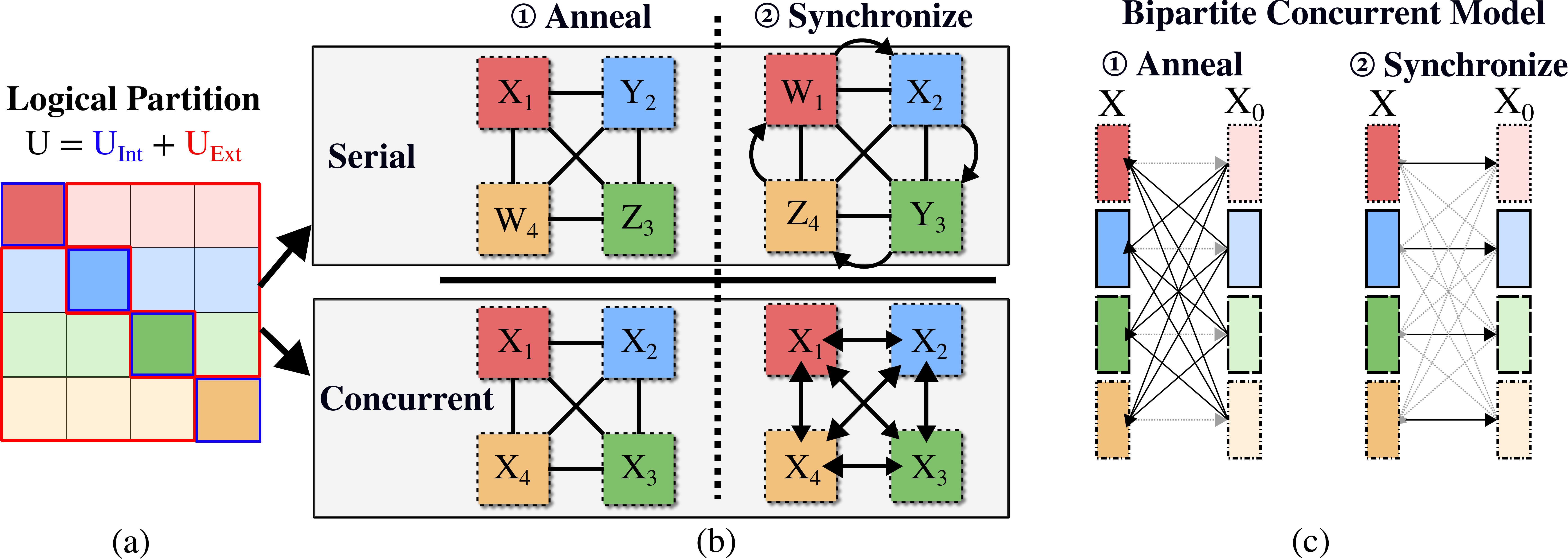}
    \caption{Illustration of the two primary parallel IM execution models. (a) A graphical illustration of a logical partition of the target potential $U$ into $U_{Int}$ and $U_{Ext}$. Here $U(X)$ is depicted as a quadratic function $U(X)=\frac{1}{2}X^TJX$ for some symmetric matrix $J$. The internal interactions comprising $U_{Int}$ have a block diagonal structure (outlined in blue) while the external interactions lie on the off-diagonal elements (highlighted in red). (b) The operation of the two execution models: serial and concurrent. During the annealing phase, serial execution simultaneously optimizes multiple independent replicas $X$, $Y$, $Z$, and $V$, while concurrent operation optimizes a single replica $X$. During a synchronization step, serial execution transfers replicas between chips to conditionally optimize the next subspace in a block Gibbs fashion and concurrent operation globally synchronizes the local spin representations of all subsystems. (c) A bipartite model representing the concurrent model. During an annealing stage, the $X$ subsystems are independent given $X_{0}$. During synchronization, $X_{0}$ is sampled from $X$, and the process begins again in the next epoch.}
    \label{fig:conc_batch_diag}
\end{figure*}
Two primary approaches have been proposed for solving out-of-capacity problems: \emph{serial} and \emph{concurrent}, each of which is illustrated in Fig.~\ref{fig:conc_batch_diag}.

\textit{Serial operation} can be likened to block Gibbs sampling, where the IM separately evolves each subsystem conditioned on the rest of the problem state
\begin{equation}
\label{eqn:serial}
    \text{Serial}
        \begin{cases}
        X_1^1\sim \pi(X_1|X_2^0,...,X_B^0)\\
        X_2^1\sim \pi(X_2|X_1^1,X_3^0,...,X_B^0)\\
        \hspace{2cm}\vdots\\
        X_B^1\sim \pi(X_B|X_1^1,X_2^1,...,X_{B-1}^1)\end{cases}.
\end{equation}
However, actual IM systems will have some history dependence/bias in their sample distributions, and therefore~\eqref{eqn:serial} is simply an idealization.

Serial methods have also been referred to as ``alternating minimization'' in optimization literature~\cite{boyd_distributed_2011,yi_alternating_2014,jain_low-rank_2013}, ``large-neighborhood local search'' (LNLS) in IM literature~\cite{booth_partitioning_nodate,mcgeoch_d-wave_2020}, and ``generalized heat-bath'' or ``partial resampling'' methods in computational physics~\cite{sokalMonteCarloMethods1997}. Prominent examples include hybrid D-Wave Leap algorithms, which alternate subproblem sampling with classical post-processing. If each sub-solver converges to its conditional distribution, serial operation guarantees asymptotic convergence to the stationary distribution~\cite{burns_provable_2024}. Serial algorithms have also achieved high solution quality in a number of combinatorial applications, particularly when combined with classical post-processing. However, serial algorithms necessarily increase latency by sacrificing parallelism afforded by dynamical system optimization.

In contrast, \textit{concurrent operation} evolves all subsystems simultaneously, conditioned on the previous problem state:
\begin{equation}\label{eqn:parallel_MC}
    \text{In Parallel}
        \begin{cases}X_1^1\sim \pi(X_1|X_2^0,...,X_B^0)\\
        X_2^1\sim \pi(X_2|X_1^0,X_3^0,...,X_B^0)\\
        \hspace{2cm}\vdots\\
        X_B^1\sim \pi(X_B|X_1^0,X_2^0,...,X_{B-1}^0)\end{cases}
\end{equation}

Parallel updates have long been utilized in computational statistical physics to accelerate simulations of classical and quantum systems. Typically, highly parallel spin updates are limited to lattice-based systems~\cite{ogielskiDynamicsThreedimensionalIsing1985,fangParallelTemperingSimulation2014}. Regular lattices enable researchers to apply specific domain decompositions which yield a factorized system~\cite{newmanMonteCarloMethods1999}. The factorized subsystems can then be sampled in parallel without any sampling bias. If the domain decomposition does not produce a factorized distribution, then the resulting sample distribution will differ (perhaps significantly) from the target $\pi$.

Spin glasses derived from combinatorial optimization problems do not have known factorizing decompositions. However, the pursuit of high-performance optimization has driven further proposals for parallel updates in spite of their inherent inaccuracies. Examples include proposed analog IM archectures~\cite{sharma_increasing_2022} and more recently in an AQC system~\cite{santos_enhancing_2024}. Notice that such parallel physical systems differ from other parallel \emph{computational} IMs~\cite{tatsumura_scaling_2021, sundara_raman_sachi_2024} in subtle but important ways. 

%In both types, problem partitioning causes a loss of efficiency in problem solving. The slowdown in computational systems is similar to traditional parallel computing: communication between units can slow down dependent computation. 

A computational IM can have multiple subunits and thus execute a parallel algorithm. Yet that parallel algorithm can still follow serial dynamics of an ideal, monolithic dynamical system~\cite{tatsumura_scaling_2021}. Physical IMs, on the other hand, are different. The dynamics are dictated by how the components are connected. While spins on the same chip can impact each other practically instantaneously, spins from different chips can not: the continuous nature of both the time and state of spins plus the limited communication bandwidth (relative to ultra-fast dynamics of physical IMs) dictate that impacts of remote spins are necessarily delayed and approximate. This approximation changes the actual Hamiltonian of the parallel machines. Slowing down the dynamics obviously can reduce approximation error, but is clearly suboptimal. In this paper, we focus on the impact of this approximation error in classical IMs. Similar effects in quantum systems have been studied in~\cite{santos_enhancing_2024} using Trotterization and BCH truncation. 

In a realistic parallel IM implementation with limited communication bandwidth, 
partitions can only exchange spin states at a frequency allowed by the communication fabric. Thus, the state of a spin from an external partition is only updated periodically and remains frozen until the next update. The partitioned spin glass problem can be described using a bipartite system model, illustrated in Fig.~\ref{fig:conc_batch_diag}(c). We distinguish between active degrees of freedom ($X(t)$) and latent spin copies $X_0\triangleq X(\lfloor \frac{t}{\tau} \rfloor \tau)$ encoding the system state at the last synchronization epoch. In practice, the latent states may either be explicitly represented as hardware copies~\cite{sharma_increasing_2022} or summarized into a linear bias term~\cite{raymond_hybrid_2023,booth_partitioning_nodate}. 

In either case, the partitioned solver acts as a block Gibbs sampler, alternating between sampling $X\sim \pi(Y|X_{0})$ and $X_{0}\sim \pi(Y|X)$. The $X$ sampling phase occurs in continuous time on each dynamical system, while sampling $X_{0}$ occurs during a discrete synchronization step. When viewed as a bipartite system, ``concurrent mode'' is an instance of partial resampling~\cite{sokalMonteCarloMethods1997} with exact conditional distributions.  

Moreover, the conditional distribution $\pi(X|X_{0})$ factorizes over partitions into $\pi(X|X_{0})=\prod_{p=1}^B \pi_b(X_p|X_{0})$, with $\pi_b(X_b|X_{0})$ being the stationary distribution of subsystem $p$. Therefore the \emph{conditional} can be exactly sampled using parallel subsystems. However, the primary problem is immediately apparent. Unless the target distribution $\pi$ permits the same factorization, concurrent mode dynamics will not have $\pi$ as the stationary target. A key aim of this work is to determine how closely the factorized model \emph{approximates} $\pi$.

\subsection{System Dynamics}
\newcommand{\ft}{{\lfloor t\rfloor}}
We consider a system with a Lyapunov function
\begin{equation}
    H(X)=-\frac{\xi}{2}U(x)+\sum_{i=1}^d f_i(x_i)
\end{equation}
where $\xi$ is a system-dependent constant, $U(x)$ is the system potential implementing pairwise interactions and $\{f_i(X):\R^N\to\R\}$ are a set of functions 
corresponding to self-interaction terms (e.g., sub-harmonic injection locking) and external fields. Notably, we assume $U$ is a \emph{separable} potential:
\begin{definition}[Separable Potential]
    A function $U:\R^N\to\R$ is a \emph{separable} potential if it can be expressed as
    \begin{equation}
        U(x)=\sum_{i\neq j}u_{ij}(x_i,x_j)
    \end{equation}
    where $\{u_{ij}\}$ is a set of continuously differentiable functions.
\end{definition}

The IM models considered here include a Brownian noise term $dW_t$ in their system dynamics to model analog noise and/or implement thermal annealing~\cite{wang_oim_2019,sharma_augmenting_2023}. We abuse notation for the sake of simplicity by denoting each vector or scalar Brownian term as $dW_t$, with the dimension clear from context.

\begin{example}[Linear Systems]
    The analog electronic arch proposed in Refs.~\cite{afoakwa_brim_2021,sharma_augmenting_2023,sharma2023combining} uses capacitor voltage as a continuous approximation of spin. The system can be modeled using the SDE
    \begin{equation}\label{eqn:sde_linear}
        dV=\frac{1}{RC}JVdt+\sqrt{2\beta(t)^{-1}}dW_t
    \end{equation}
    where $V\in \R^N$ describes the voltages across $N$ resistively-coupled capacitors with system time constant $RC$, $J$ is a symmetric coupling matrix, and $\sqrt{2\beta(t)^{-1}}$ is a time-dependent drift term. In practice there are added potential terms confining $V$ to $[GND, VDD]$, however Eqn.~\ref{eqn:sde_linear} is a useful approximation within the normal operating region. The Hamiltonian is correspondingly:
    \begin{equation}
        H(V)=-\frac{1}{2RC}V^TJV
    \end{equation}
    which is linear and therefore separable. For numerical simulations we set $C=\SI{50}{fF}$ and $R=\SI{310}{k\Omega}$ and integrate~\eqref{eqn:sde_linear} using an Euler-Maruyama discretization and a 10.
    
\end{example}
\begin{example}[Kuramoto Model]
    A network of $N$ Kuramoto oscillators with phases $\{\theta_i\}_{i=1}^N$ with a sub-harmonic injection locking (SHIL) term obey a system of coupled SDEs~\cite{wang_oim_2019}
    \begin{equation}
    \begin{split}
        d\theta_i=-K_J(t)\sum_{j\neq i}J_{ij}\sin(\theta_i-\theta_j)dt\\
        +K_S(t)\sin(2\theta_i)dt+\sqrt{2\beta(t)^{-1}}dW_t
    \end{split}
    \end{equation}
    where $K_J(t)$ and $K_S(t)$ are time-dependent coefficients defining the energy landscape and $J_{ij}$ are the elements of a symmetric coupling matrix. The corresponding Hamiltonian is:
    \begin{equation}H(\theta)=-K_J(t)\sum_{j\neq i}J_{ij}\cos(\theta_i-\theta_j)\\
        +\frac{K_S(t)}{2}\sin(2\theta_i)dt
    \end{equation}
    which is obviously separable.
\end{example}

Note that this framework does not handle nonlinear activation functions as proposed in Ref.~\cite{bohm_order--magnitude_2021}, indicating a direction for future work.

By restricting ourselves to separable systems, we can perform block decompositions $X=(X_1,\dots,X_B)$ with separable gradients
\begin{equation}
    \pdv{U}{x_i}=\sum_{p=1}^B\sum_{x_j\in X_p}\pdv{u_{ij}}{x_i}
\end{equation}
allowing us to express the system SDE
%\begin{widetext}
\begin{equation}\label{eqn:full_system}
\begin{split}
    dX_p=\left(\overbrace{\nabla_{X_p}\left[\sum_{x_i\in X_p}f_i(x_i)+\sum_{\substack{i\neq j\\x_i,x_j\in X_p}}u_{ij}(x_i,x_j)\right]}^{\nabla_{X_p}U_{Int}(X(t))}\right.
    \\\left.+\overbrace{\nabla_{X_p}\left[\sum_{\substack{x_i\in X_p\\x_j\not\in X_p}}u_{ij}(x_i,x_j)\right]}^{\nabla_{X_p}U_{Ext}(X(t)))} \vphantom{}\right) dt + \sqrt{2\beta^{-1}(t)} dW_t.
\end{split}
\end{equation}

%\end{widetext}
We define $\nabla_{X_p}U_{Int}(X(t))$ as ``Internal'' interactions occurring within the variable block and $\nabla_{X_p}U_{Ext}(X(t))$ as ``External'' interactions occurring between distinct blocks. Viewing the system as a network of interacting nodes, the ``External'' interactions are the set of ``cut'' edges formed by the partition. 

Suppose that we condition the dynamics of block $X_p$ on all external blocks by fixing their states at time $s \leq t $. Then the current
\begin{equation}
    \nabla_{X_p}U_{Ext}(X_p(t), X_{\setminus p}(s))
\end{equation}
can be regarded as either a fixed external field (for the linear system) or a self-interaction term (in the case of the Kuramoto model), conditioned on the states of other blocks. 

Then the system dynamics for each block are given by:
\begin{equation}\label{eqn:time_delay}
\begin{split}
    dX_p&=\left[\nabla_{X_p}U_{Int}(X_p(t))+\nabla_{X_p}U_{Ext}(X_p(t), X_{\setminus p}(s))\right]dt\\& + \sqrt{2\beta^{-1}} dW_t
\end{split}
\end{equation}

\begin{comment}
In even further generality, we can assume a sparsely connected network of Ising machines with a maximum hop distance $d_{\max}$ given by
\begin{equation}\label{eqn:time_delay}
\begin{split}
    dX_i=&\left[\nabla_{X_i}U_{Int}(X(t))\right.\\&+\left.\nabla_{X_i}U_{Ext}(X_i(t), X_\excl{i}_{prev})\right]dt + D dW_{k,t}
\end{split}
\end{equation}
where \[X_{\excl{i},{prev}}=(X_{j}(\left\lfloor \frac{t-\tau d(i,j)}{\tau}\right\rfloor\tau))\] stores the most up-to-date information propagated from other states in the network.
\end{comment}

We define the \emph{synchronization epoch} as $\tau=\max(t-s)$. Clearly, Eqn.~\eqref{eqn:time_delay} recovers Eqn.~\ref{eqn:full_system} in the $\tau\to0$ (infinite bandwidth) limit, however with finite bandwidth fundamentally incurs some approximation. For a given realization of $X_{0}\triangleq X(s)$, the dynamics described by Eqn.~\eqref{eqn:time_delay} will in general have different fixed points than~\eqref{eqn:full_system}. Taking the example of a deterministic linear model, we have

\begin{equation}\label{eqn:approx_lin_sde}
    dX=J_{Int}X(t)dt+J_{Ext}X_{0}dt.
\end{equation}

% which has the Lyapunov function

% \begin{equation}\label{fixed_points}
%     H_t(x)=-\frac{1}{2}X(t)^TJ_{Int}X(t) - X_{0}^TJ_{Ext} X(t)
% \end{equation}

For sufficiently small synchronization epochs $\tau$, ~\eqref{eqn:approx_lin_sde} is a good approximation to the full system dynamics. However, as $\tau$ increases we have increasing divergence from the ideal system dynamics. The second term provides a linear bias aligning $X(t)$ parallel to $J_{Ext}X_{0}$. When the spectral radius of $J_{Ext}$ is significantly greater than $J_{Int}$, then the external linear term can become large enough to dominate the time evolution of $X_t$. Furthermore, when the maximum-magnitude eigenvalue of $J_{Ext}$ is negative, sufficiently large values of $\tau$ result in oscillatory behavior between epochs, as we now demonstrate.

\begin{figure}
    \centering
    \begin{subfigure}[t]{1.0\linewidth}
    \includegraphics[width=0.8\linewidth]{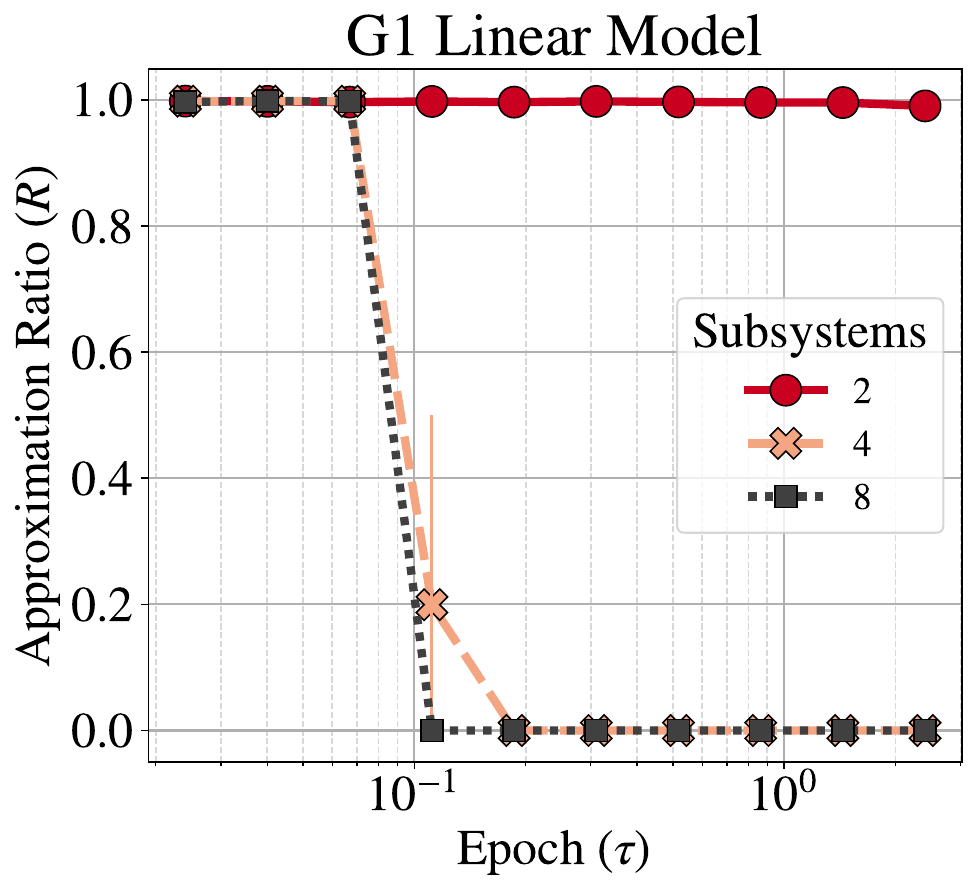}
    \subcaption{}\label{fig:g1_div}
    \end{subfigure}
    
    \begin{subfigure}[t]{1.0\linewidth}
    \includegraphics[width=0.8\linewidth]{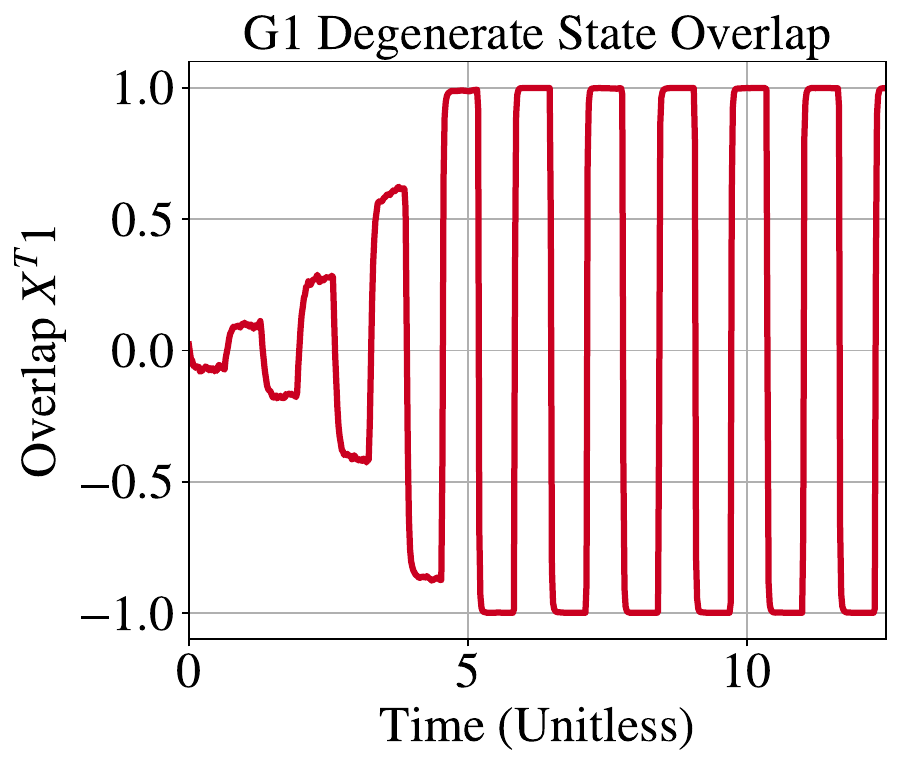}
    \subcaption{}\label{fig:g1_osc}
    \end{subfigure}
    \caption{Simulated linear model behavior on graph G1 versus unitless time (a) Concurrent mode cut value versus synchronization epoch $\tau$ for the unweighted graph G1 using 2, 4, and 8 subsystems. Cut value 0 indicates that the system diverged to the degenerate $\pm 1^N$ state. (b) State overlap $\langle X, 1^N\rangle$ versus time for 4 subsystem concurrent execution with a $\tau=0.6$ synchronization epoch.}
    \label{fig:erdos-renyi-divergence}
\end{figure}

Fig.~\ref{fig:erdos-renyi-divergence} shows the behavior of a linear process with increasing synchronization period $\tau$ on the GSet~\cite{noauthor_index_nodate} benchmark graph G1. Fig.~\ref{fig:g1_div} shows the approximation ratio $R$ for several block counts $B$ with increasing synchonization epoch. While the system performs well with large $\tau$ with $B=2$, increasing the number of blocks results in a sharp decline in solution quality. The graph has purely anti-ferromagnetic $-1$ couplings, resulting in a dominant eigenvector with strictly positive elements and negative eigenvalue for both $J_{Int}$ and $J_{Ext}$. 

For $B=2$, the spectral radii $\rho(J_{Int})$ and $\rho(J_{Ext})$ are comparable (25.1 and 24.9 respectively). Increasing to $B=4$ results in $J_{Int}$ being dominated by $J_{Ext}$ ($\rho(J_{Int})=13.1,\rho(J_{Ext})=37.0$), with $B=8$ further exacerbating the difference ($\rho(J_{Int})=7.5,\rho(J_{Ext})=43.0$). The minimum energy state of the approximate Lyapunov function within the hypercube $[-1,1]^N$ then becomes the degenerate $\{-1\}$/$\{+1\}$ states, depending on $X_0$. For a random initialization, we typically have $X_0^T \bm{1}>0$. The dominant eigenvector of $J_{Ext}$ then tends to amplify the magnitude of $X(t)^T \bm{1}$ over time, as shown in  Fig.~\ref{fig:g1_osc}. 

As a result, the system oscillates between the two poles on each synchronization epoch. These oscillations are purely an artifact of using a bounded $X\in[-1,1]^N$ domain, the true problem is that \emph{the factorized stationary distribution no longer faithfully approximates the highly coupled target distribution}.

However, we find that this behavior is limited to fully anti-ferromagnetic models with dominant spectral modes, and we do not observe this behavior in benchmarks with mixed ferromagnetic and anti-ferromagnetic couplings or in problems with non-unit weights. Nevertheless, parallel execution models still introduce limits on convergence, as shown theoretically and numerically in Sections~\ref{sec:convergence} and~\ref{sec:experiments}.

\section{Convergence Properties}\label{sec:convergence}
In this section we discuss the weak convergence properties of the serial and concurrent execution models. We begin by briefly reviewing existing results for serial execution models, which are more well-studied in literature. We then present novel results concerning the weak convergence of concurrent execution models. We first demonstrate finite lower bounds for asymptotic KL-divergence between the sample and target distributions, then provide sufficient conditions for contraction in 1-Wasserstein distance in each epoch.

The latter results are not \textit{directly} usable, as Wasserstein distances are currently intractable to estimate for general probability distributions~\cite{chewiStatisticalOptimalTransport2025}. However, they will provide insight into the behavior of the concurrent processes from which we formulate empirically usable heuristics in later sections. Moreover, our results may be of independent interest in stochastic thermodynamics, where Wasserstein distances have been recently applied~\cite{nakazato_geometrical_2021,dechant_minimum_2022,dechant_thermodynamic_2019}.

\subsection{Serial Execution}
As explored in Refs.~\cite{ding_langevin_2021, raymond_hybrid_2023, burns_provable_2024}, conditional sampling leaves the stationary distribution invariant. Therefore, we have no finite convergence lower bound for a serial execution model (Eqn.~\eqref{eqn:serial}). Note, however, that the \emph{rate} of convergence may decay significantly. 

For instance, the (relatively) well-behaved functions examined in Ref~\cite{burns_provable_2024} incurred a linear slowdown in $\kl{\mu}{\pi}$ convergence. Spin glass optimization is far from well-behaved. We strongly conjecture that conditional sampling on highly non-convex potentials incurs super-polynomial slowdowns, though we leave exact characterization to future work.

\subsection{Concurrent Execution}
\subsubsection{Lower Bounds}
It is clear that the concurrent execution model only approximates full Ising machine behavior. One method of comparing the approximation accuracy is to consider the difference between the sample distribution of the concurrent IM network and that of a monolithic IM. The distance (or divergence) between these distributions (Sec.~\ref{sec:prelims_dist}) in the limit $t\to \infty$ is the \emph{bias} of the approximation.

We consider two processes which approximate real-world machine construction. For clarity, we denote the state of the first process as $Y(t)\triangleq Y_t$ and the second as $X(t)\triangleq X_t$.

The first is an ideal overdamped Langevin process $Y_t$ with instantaneous interaction between components:
\begin{equation}\label{eqn:ideal_process}
    dY_t = \big(\nabla U_{Int}(Y_t) +\nabla U_{Ext}(Y_0)\big)dt + \sqrt{2\beta^{-1}}dW_t.
\end{equation}
This is an approximation of a "monolithic" physical Ising machine, where the on-chip spin interaction is fast enough to be modeled as instantaneous. The second process models multiple independent chips with a finite communication bandwidth. In such a system, practical implementation may not allow all spin state changes on one chip to be seen by other chips (let alone instantaneously). One possibility is the periodic exchange of state updates among chips~\cite{sharma_increasing_2022}. We approximate such a design with a concurrent-mode process $X(t)\triangleq X_t$
\begin{equation}
    dX_t = \big(\nabla U_{Int}(X_t) +\nabla U_{Ext}(X_0)\big)dt + \sqrt{2\beta^{-1}}dW_t,
\end{equation}
where $t \in [0, \tau)$ and $X_0$ is the system state at the last synchronization time point. $X_\tau$ is a Markov process obeying Eqn.~\eqref{eqn:parallel_MC}, thereby fitting the ``concurrent execution'' model defined in the previous section. By definition, the processes $X_t$ and $Y_t$ are synchronously coupled, as they share the same Brownian noise source. The $X_t$ process represents the networked analog system, while $Y_t$ gives the dynamics in the $\tau\to 0$ limit. Let $\nu_t$ and $\mu_t$ be the distribution underlying processes $Y_t$ and $X_t$ respectively at time $t$.

Given suitable regularity conditions on $U$ (satisfied by a relaxed Ising Hamiltonian), $\nu_t$~\eqref{eqn:ideal_process} is guaranteed to converge to the stationary distribution $\pi$ (albeit in time exponential in the problem dimension)~\cite{cheng_sharp_2020}. However, the same is \emph{not} true for $\mu_t$, it is asymptotically biased. We can intuitively see why: $\mu_t$ factorizes across partitions, whereas $\pi$ (in general) does not. However, we would still like to know \emph{how close} the two distributions are, and what parameters affect the asymptotic bias.

Suppose we start both processes from the same initial point $X_0=Y_0$ and then observe the difference between their sample paths. Recall from Section~\ref{sec:prelims} that the KL-divergence (AKA relative entropy) is given by
\begin{equation}
    \kl{\mu}{\nu}=\int_\region\mu(x)\log\frac{d\mu}{d\nu}(x)dx.
\end{equation}

We can express the relative entropy between the concurrent and ideal systems using the gradient error using the following proposition (proved in Appendix~\ref{appdx:kl_proof}).

\begin{proposition}[KL-Divergence of the Approximated Process]\label{prop:kl}
    Let $\mu_\tau$ and $\nu_\tau$ be the probability distributions of the concurrent process and full system processes respectively at time $\tau \geq 0$. Suppose \begin{equation}
       \tau\langle\| \delta\nabla U\|^2\rangle\!\triangleq\!\!\int_0^\tau\!\!\norm{\nabla U_{Ext}(X_t)\!-\!\nabla U_{Ext}(X_0)}^2dt
    \end{equation}
    is the accumulated gradient error from process $X$ up to time $\tau$.
    Then the KL-Divergence between the two measures is
    \begin{equation}\label{eqn:kl_div_result}
        \kl{\mu_\tau}{\nu_\tau}=\frac{\beta}{4}\mathbb{E}_{\mu_\tau}[\tau\langle\| \delta\nabla U\|^2\rangle]
    \end{equation}
\end{proposition}

Eqn.~\eqref{eqn:kl_div_result} makes intuitive sense: the distance between our ``ideal'' process $Y_t$ and our approximated process $(X_t)$ is a function of the accumulated gradient error. This can also be understood using existing notions of information flow from the literature. The Liang-Kleeman information flow~\cite{liang_information_2016}, which measures the effect of one subsystem on the entropy of another, also depends on instantaneous partial derivatives of the potential function w.r.t. subsystem parameters. By conditioning the external parameters, we impose a bottleneck on system information exchange, thereby limiting the change in the entropy of each subsystem. 

Similarly, the relative Fisher information $FI(\mu_\tau\Vert\nu_\tau)$ between $\mu_\tau$ and $\nu_\tau$ can be expressed as
\begin{equation}
    \textrm{FI}(\mu_\tau\Vert\nu_\tau)=\mathbb{E}_{\mu_\tau}[\norm{\tau\langle(\nabla^2\delta  U)\delta\nabla U\rangle}^2]
\end{equation}
where \[
\nabla^2\delta  U=\nabla^2 U_{Ext}(X_s)-\nabla^2 U_{Ext}(X_0),\]
using the notation $\nabla^2f(x)$ as the Hessian of $f(x)$. The Fisher information captures the local differential structure affecting the differences between the measures. Therefore, our ``gradient error'' framework also captures the local curvature of measure space. A corollary is that linear models obeying Eqn.~\eqref{eqn:sde_linear} have no curvature in measure space, while non-linear Ising oscillator-based Ising machines navigate a richer landscape. We leave a full comparison between linear and non-linear models to future work.

Note that by Pinsker's inequality we can restate~\eqref{eqn:kl_div_result} as an upper bound on the more familiar total variation distance $\delta$
\begin{equation}
    \delta(\mu,\nu)\leq \sqrt{\frac{\beta}{8}\mathbb{E}_{\mu_\tau}[{\tau\langle\norm{\delta\nabla U}^2\rangle}]}.
\end{equation}

\subsubsection{Sufficient Conditions for Convergence}

Eqn.~\eqref{eqn:kl_div_result} provides useful theoretical lower bounds, however it does not provide much direct insight into problem parameters which may impact convergence, or when convergence to $\pi$ will stall. In this section we use concepts from optimal transport~\cite{villaniTopicsOptimalTransportation2003} to bound the provide \emph{sufficient} conditions for the concurrent process to decrease the distance to $\pi$ within an iteration.

Recall from Sec.~\ref{sec:prelims} that the 1-Wasserstein distance is given by
\begin{equation}
    W_1(\mu_t,\nu_t)=\inf_{\gamma\in\mathcal{C}(\mu_t,\nu_t)}\mathbb{E}_\gamma\left[\onenorm{X_t-Y_t}\right].
\end{equation}
The 1-Wasserstein distance satisfies the triangle inequality 
\begin{equation}
    W_1(\mu_t,\pi)\leq W_1(\pi,\nu_t)+W_1(\nu_t,\mu_t)
\end{equation}
where $\pi$ is our target distribution.

For the ideal Langevin process, $W_1(\pi,\nu_t)\leq C(t)W_1(\pi,\nu_0)$, where $C(t):\R\to(0,1)$ is a decreasing function of $t$~\cite{cheng_sharp_2020,chewi_analysis_2022,vempala_rapid_2019,dalalyan_user-friendly_2019}. For general non-convex problems, convergence rates are often logarithmic (see Ref.~\cite{chiang_diffusion_1987} for convergence in total variation distance, for instance), making convergence essentially asymptotic. Since $\mu_0=\nu_0$, if we have
\begin{equation}
\begin{split}
    W_1(\mu_t,\nu_t)+C(t)W_1(\pi,\mu_0)&< W(\pi,\mu_0)\\
    \frac{W_1(\mu_t,\nu_t)}{1-C(t)}&< W(\pi,\mu_0)
\end{split}
\end{equation}
then we have some guaranteed contraction
\begin{equation}
 W_1(\pi,\mu_t)<  W(\pi,\mu_0).
\end{equation}
The key, then, is to estimate $W_1(\mu_t,\nu_t)$. For any functional $G(\mu_t,\nu_t)\geq W_1(\mu_t,\nu_t)$, we can take the stronger condition
\begin{equation}\label{eqn:template}
    \frac{G(\mu_t,\nu_t)}{1-C(t)}< W(\pi,\mu_0)
\end{equation}
as sufficient for contraction in $W_1$. In this section, we provide two such functionals upper bounding $W_1(\mu_t,\nu_t)$, both proved in Appendix~\ref{appdx:w1_proof}.

As in the previous section, the processes $X_t$ (concurrent) and $Y_t$ (ideal) are synchronously coupled and begin from the same initial state. First suppose we have access to both processes. Then we can upper bound $W_1(\mu_t,\nu_t)$ using the following result:
\begin{proposition}[Upper Bound (I)]\label{prop:w1_1}
    Let $\mu_t$, $\nu_t$ be the distribution of $X_t$ and $Y_t$ within a concurrent execution epoch. Then
    \begin{equation}\label{eqn:upper_bound_1}
\begin{split}
    W_1(\mu_t,\nu_t)\leq&\mathbb{E}_{\nu_t\times\mu_t}\lVert\nabla U(Y_t)-\nabla U(X_t) \rVert_1\\&+\mathbb{E}_{\mu_t}\lVert\nabla U_{Ext}(X_0)-\nabla U_{Ext}(X_t)\rVert_1.
\end{split}
\end{equation}
\end{proposition}
Substituting~\eqref{eqn:upper_bound_1} into~\eqref{eqn:template} we obtain the contraction condition for $W_1$:
\begin{corollary}
Let $\mu_t$ be the distribution of $X_t$ within a concurrent execution epoch. Then the condition
    \begin{equation}\label{eqn:w1_convergence}
\begin{split}
    &(1-C(t))^{-1}\left[\mathbb{E}_{\nu_t\times\mu_t}\lVert\nabla U(Y_t)-\nabla U(X_t) \rVert_1\right.\\
    &\hspace{1.05cm}\left.+\mathbb{E}_{\mu_t}\lVert\nabla U_{Ext}(X_0)-\nabla U_{Ext}(X_t) \rVert_1\vphantom{}\right]\\
    &<W_1(\mu_0,\pi)
\end{split}
\end{equation}
is sufficient to guarantee $W_1(\mu_t,\pi)<  W(\mu_0, \pi)$.
\end{corollary}
However, outside of simulations we do not have access to a synchronously coupled, ideal Langevin process. With some additional assumptions, we can express a contraction condition that only depends on problem structure and the trajectory $X_t$. 

Let $\cal{D}$ be the domain of the Hamiltonian $H$ (e.g., the unit hypercube for the linear model in Eqn.~\eqref{eqn:sde_linear}). We assume that $\nabla U(X)$ is $L$-Lipschitz continuous in $\cal{D}$, that is, there exists a constant $L\geq 0$ such that for all $X,Y\in \cal{D}$
\begin{equation}
    \begin{split}
        \norm{\nabla U(X)-\nabla U(Y)}_p&\leq L\norm{X-Y}_p
    \end{split}
\end{equation}
for $p\in\{1,2\}$, where we use the same constants for the $\ell_1$ and $\ell_2$ norms for simplicity.  We only assume Lipschitz continuity on $\cal{D}$, allowing for the possibility of non-globally Lipschitz confining potentials, such as cubic or quartic polynomials, if $\cal{D}$ is compact. In all analysis we therefore assume that the processes $X_t$, $Y_t$ are confined to $\cal{D}$.

\begin{proposition}[Sufficient Condition for Convergence (II)]\label{prop:w1_2}
    Let $\mu_t$, $\nu_t$ be the distribution of $X_t$ and $Y_t$ within a concurrent execution epoch. Assume that $0\leq tL<1$ and that $\mathbb{E}_\gamma\onenorm{X_t-Y_t}$ is non-decreasing with time, where $\gamma$ is the optimal coupling between $\mu_t$ and $\nu_t$. Then
    \begin{equation}\label{eqn:upper_bound_2}
\begin{split}
    W_1(\mu_t,\nu_t)\leq \frac{\mathbb{E}_{\mu_t}\onenorm{\nabla U_{Ext}(X_0)-\nabla U_{Ext}(X_t)}}{1-tL}
\end{split}
\end{equation}
is sufficient to guarantee $W_1(\mu_t,\pi)<  W(\mu_0, \pi)$.
\end{proposition}
As before, we immediately obtain a contraction condition for $W_1(\mu_t,\pi)$.
\begin{corollary}
Let $\mu_t$ be the distribution of $X_t$ within a concurrent execution epoch. Then the condition
    \begin{equation}\label{eqn:w1_convergence_ii}
\begin{split}
    &(1-C(t))^{-1}\frac{\mathbb{E}_{\mu_t}\onenorm{\nabla U_{Ext}(X_0)-\nabla U_{Ext}(X_t)}}{1-tL}\\
    &<W_1(\mu_0,\pi)
\end{split}
\end{equation}
is sufficient to guarantee $W_1(\mu_t,\pi)<  W(\mu_0, \pi)$.
\end{corollary}

To numerically evaluate our bounds, we simulated ideal $Y_t$ and concurrent $X_t$ processes on a 12-spin ferromagnetic lattice with 2, 4, and 6 subsystems with fixed inverse temperature $\beta = 10$. For computational simplicity we focus on the linear IM model (Equation~\eqref{eqn:sde_linear}). For each process, we compute a discrete empirical distribution and compare to the true Gibbs distribution. Appendix~\ref{sec:appdx_experimental} contains further experimental details.

\begin{figure}[h]
    \centering
    \includegraphics[width=0.9\linewidth]{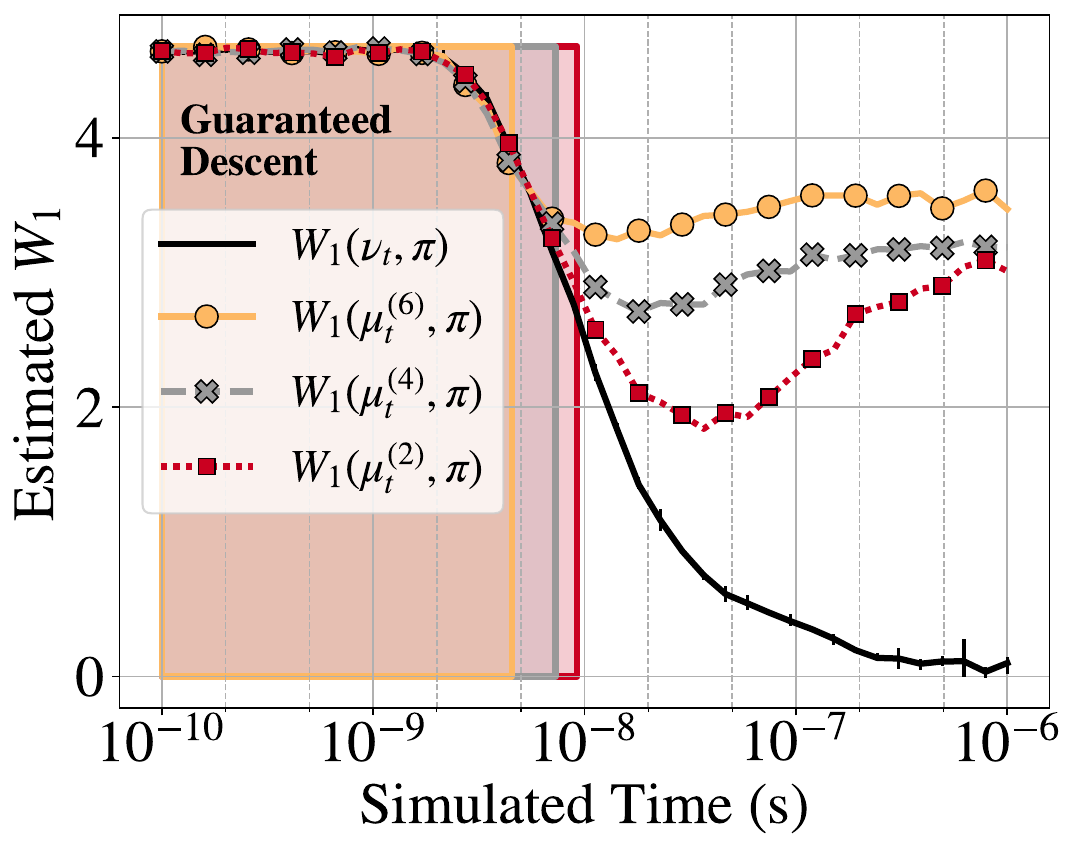}
    \caption{Simulated $W_1$ convergence on a 12-spin Ising lattice for a parallel, concurrent process and an ideal Langevin process. The number of subsystems $B$ for the concurrent process was either 2, 4, or 6, and is denoted by the superscript $\mu^{(B)}_t$. Each solver started from a uniform random distribution. Error bars show the effect of sampling error (primarily visible on the $W_1(\nu_t,\pi)$ curve).}
    \label{fig:wasserstein}
\end{figure}

Fig.~\ref{fig:wasserstein} shows the estimated $W_1$ for the concurrent processes (with the subsystem count $B$ denoted by the superscript $\mu^{(B)}_t$) compared against the ideal processes. For $\tau < 10^{-8}$, the concurrent processes are approximately equal to the ideal $Y_t$ process in Wasserstein distance. However, past that point the processes begin to deviate. As expected, the ideal process approximately converges to the true distribution. In contrast, the concurrent processes have a nonzero minimal Wasserstein distance, after which $W_1(\mu_t^{(B)}, \pi)$ begins increasing.

We also compute $\mathbb{E}_{\nu_t\times\mu_t}\lVert\nabla U(Y_t)-\nabla U(X_t) \rVert_1$ and $\mathbb{E}_{\mu_t}\lVert\nabla U_{Ext}(X_0)-\nabla U_{Ext}(X_t) \rVert_1$ during the simulation process. With these values, we compute the maximum $\tau$ with guaranteed $W_1$ contraction, shown as the shaded regions in Fig.~\ref{fig:wasserstein} (with different vertical lines marking the region boundaries, color-matched to the corresponding curves). While the $\tau$ upper bounds are conservative, we note that they are within an order of magnitude of the minimal $W_1$ for each block count. Bound II results (not shown) were overly conservative, and were not satisfied beyond $10^{-9}$.

\begin{figure}[h]
    \centering
    \includegraphics[width=0.9\linewidth]{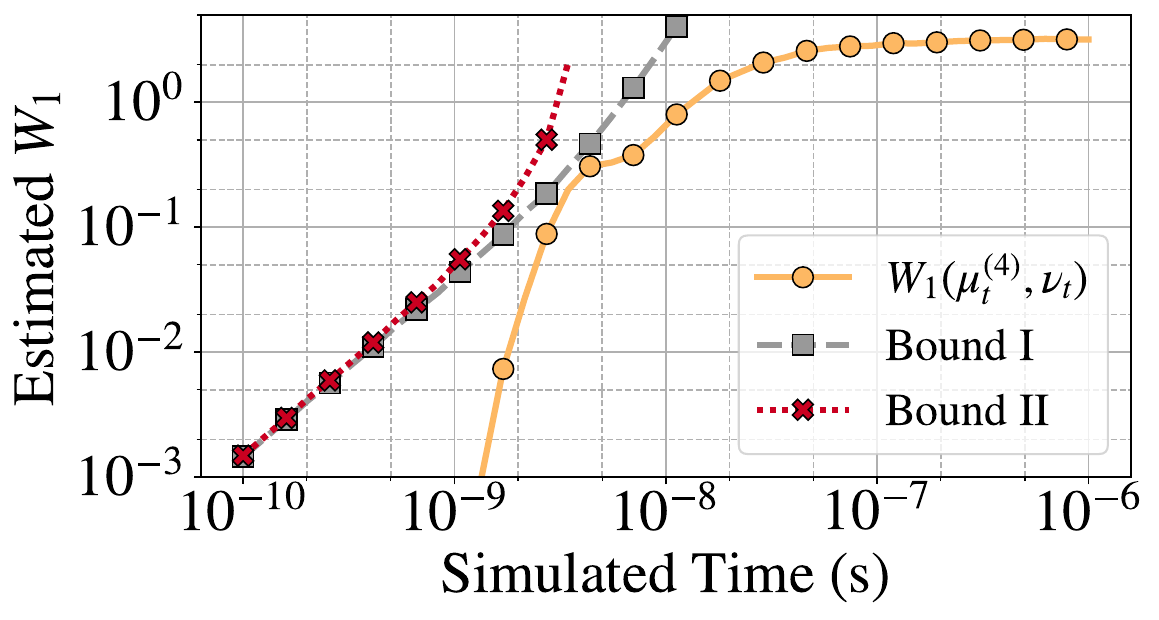}
    \caption{Comparison between the two upper bounds derived in the main text (``Bound I'' and ``Bound II'') and the estimated distance $W_1(\mu^{(4)}_t, \nu_t)$ for the $B=4$ process.}
    \label{fig:upper_bound_comparison}
\end{figure}

Fig.~\ref{fig:upper_bound_comparison} compares Bounds I and II against the computed value $W_1(\mu_t, \nu_t)$. The behavior was similar across block counts, therefore we only show the $B=4$ case for clarity. Each bound is relatively tight for $\tau<10^{-9}$, with each bound able to guarantee some contraction. However, Bound II is no longer valid for $\tau > \frac{1}{L}$ (approximately $\SI{3.9e-9}{s}$), and Bound I becomes too loose for practical use past $10^{-8}$. The latter value also roughly corresponds to the region where we observe energy divergence in larger problems, as we now show.

To demonstrate the effects of increasing $\tau$ on larger-scale problems, we compare the concurrent and serial execution models on $N=2000$ Sherrington-Kirkpatrick (SK) spin glasses~\cite{panchenko_sherrington-kirkpatrick_2012}. Each instance has the Hamiltonian
\begin{equation}
    H(s)=\sum_{i<j}J_{ij}s_is_j
\end{equation}
where $J_{ij}$ are normally distributed with variance $\frac{1}{N}$. We simulate system behavior for $\SI{20}{\mu s}$ across 3 problems with 50 trials each. Due to the lack of crosse-partition communication, the $B=1$ configuration produced consistent results across $\tau$ values. Therefore, we report the $u$ relative to the mean $B=1$ energy-per-spin ($\overline{u}_1\sim -0.756$) averaged across all trials.

\begin{figure}[h]
    \centering
    \includegraphics[width=\linewidth]{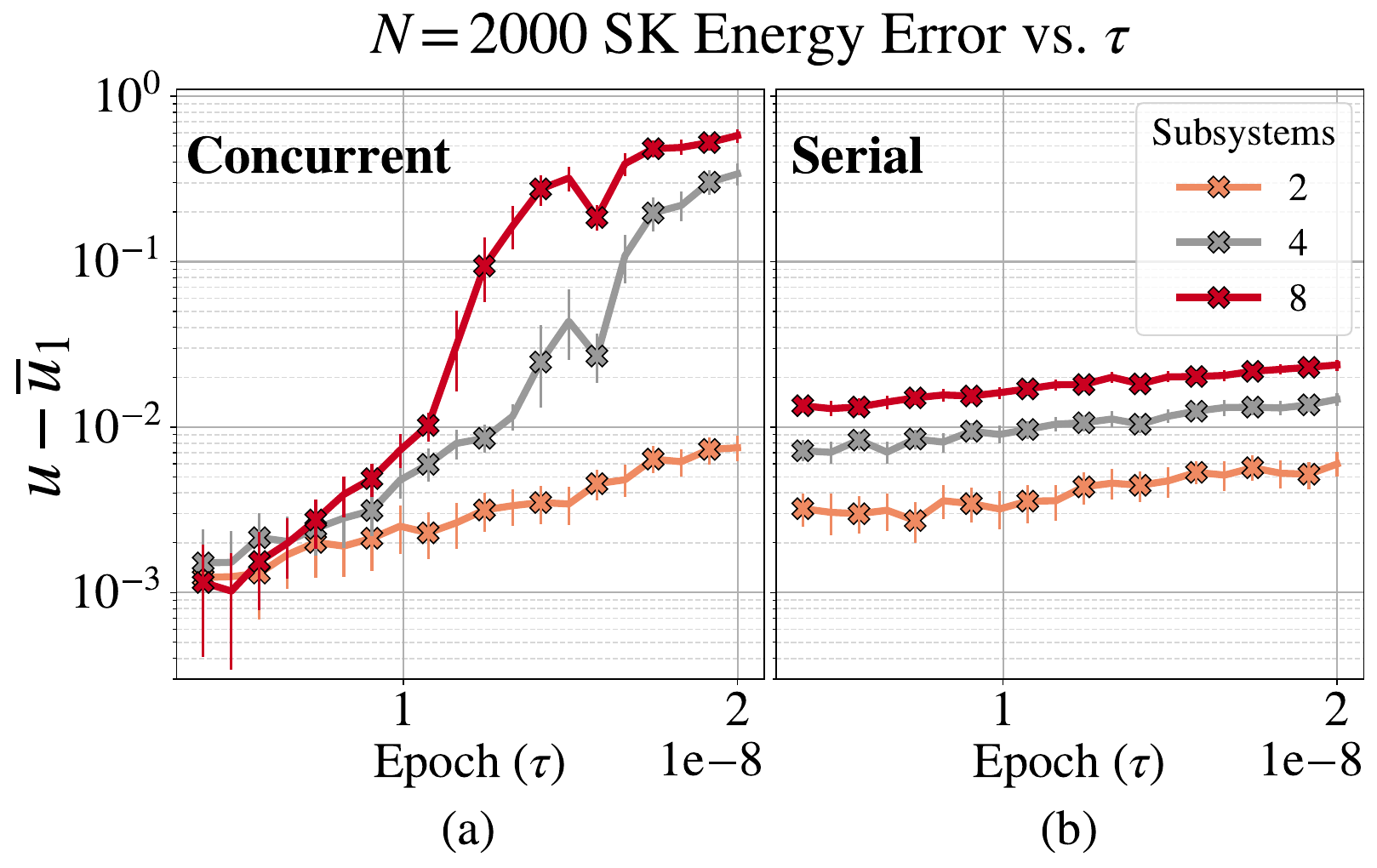}
    \caption{Deviation of the estimated energy-per-spin relative to $B=1$ case for 2000-node SK spin glass instances. Each data point represents 3 different SK instances with 50 trials each.}
    \label{fig:sk_energy_error}
\end{figure}
Fig.~\ref{fig:sk_energy_error} shows the ``energy error'' $u-\overline{u}_1$ for the serial and concurrent processes with varying block configurations and increasing epochs. For $\tau<10^{-8}$, the concurrent processes attain a consistently lower energy error than the serial processes with the same total annealing time. For higher $\tau$ values, the energy estimates begin to diverge from the monolithic system results. As in the toy ferromagnetic problem, configurations with more subsystems diverge more rapidly, with the estimated energy becoming positive for the 4 and 8 system configurations. The $B=2$ system remains close to the serial estimate, however it still demonstrates a significant loss in accuracy with increasing $\tau$. 

In contrast, the serial processes experience near-linear degradation in the estimated energy with increasing $\tau$. Unlike the concurrent system, where $\tau$ fundamentally alters system dynamics, the serial system simply requires more iterations to regain accuracy, as observed in Ref.~\cite{sharma_increasing_2022}. Serial execution models are therefore more robust to low-synchronization frequency environments. We therefore conjecture that serial models are also more robust in the presence of asynchronous updates, though we leave that comparison to future work.
\section{Empirical Analysis}\label{sec:experiments}
Having established theoretical bounds for general separable dynamical systems, we narrow our focus to linear Ising machines evolving according to Eqn.~\ref{eqn:sde_linear}. We begin by deriving a heuristic parameter choice based on Proposition~\ref{prop:w1_2}, then empirically evaluate the time and energy efficiency of serial and concurrent execution models for unconstrained binary optimization. We conclude by offering a decision tree heuristic informed by our findings.

\subsection{Heuristic Parameter Selection}
While we numerically show that Eqns.~\eqref{eqn:w1_convergence} and~\eqref{eqn:w1_convergence_ii} hold in toy problems, real-world users lack access to the $W_1$ contraction factor $C(t)$ as well as the starting $W_1(\mu_0,\pi)$. In contrast, we do have access to the problem structure in the coupling matrix $J$. Proposition~\ref{prop:w1_2} suggests that the spectrum of $J$ is a limiting factor in convergence. In this section we heuristically derive a similar relation for Ising machines by analyzing the flip time of a single spin.

Consider a single degree of freedom $x_i\in [-1,1]$ in subsystem $X_p$. We have
\begin{equation}
    \begin{split}
        dx_i = \left[\sum_{x_j \in X_{p}}J_{ij} x_j+\sum_{\mathclap{x_j \in X_{\setminus p}}}J_{ij} x_j\right]dt + \sqrt{2T}dW_t.
    \end{split}
\end{equation}
By Itô's lemma, we have that
\begin{equation}
    \begin{split}
        d(x_i^2) = \left[2x_i\cdot\left(\sum_{\mathclap{\quad x_j \in X_{p}}}J_{ij} x_j+\sum_{\mathclap{x_j \in X_{\setminus p}}}J_{ij} x_j\right) + 2T\right] dt\\
        + 2x_i\sqrt{2T}dW_t,
    \end{split}
\end{equation}
hence 
\begin{equation}
    \begin{split}
        &\E[x_{i,\tau}^2-x_{i,0}^2] =\\ &\int_0^\tau \left[\E
        \Big(\!2x_i\sum_{\mathclap{x_j \in X_{p}}}J_{ij}x_j + 2x_i\sum_{\mathclap{x_j \in X_{\setminus p}}}J_{ij} x_j\vphantom{}\!\Big)\!+\! 2T\right]\!dt.
    \end{split}
\end{equation}
Assuming that the gradient
\begin{equation}
J_i\triangleq \sum_{x_j \in X_{p}}J_{ij}x_j+\sum_{x_j \in X_{\setminus p}}J_{ij} x_j
\end{equation}
remains (relatively) constant during the annealing epoch, we have
\begin{equation}
    \E[x_{i,\tau}^2-x_{i,0}^2]\sim \E[(x_{i,\tau}^2-x_{i,0}^2)J_i+2T]\tau
\end{equation}
For combinatorial problems, the quantity of interest between $x_{i,\tau}$ and $x_{i,0}$ is the \emph{Hamming distance}~\footnote{We define the Hamming distance $d_{H}(x, y)=0$ if $\text{sign}(x)= \text{sign}(y)$ and $1$ otherwise.}, not the $\ell_1$ or $\ell_2$ distances. Therefore, we can focus our attention on \emph{spin flips} rather than the entire trajectory. Assuming that the spin flips within the epoch ($\E[x_{i,\tau}^2]=0$) and the spins began the epoch in a bifurcated state ($\E[x_{i,\tau}^2]=1$), we can solve for $\tau_{Flip}$ as
\begin{equation}
    \tau_{Flip}\sim \frac{1}{(J_i-2T)}.
\end{equation}
In the low-temperature limit $J_i >> 2T$, we further approximate as
\begin{equation}~\label{eqn:tau_flip}
    \tau_{Flip}\sim \frac{1}{J_i}\geq \frac{1}{\rho(J)}.
\end{equation}
where $\rho(J)=\max_{i}|\lambda_i|$ is the spectral radius of the coupling matrix $J$. Selecting $\tau<\tau_{Flip}$ results in relatively few spin flips occurring each epoch, decreasing the Hamming distance between $X_0$ and $X_\tau$. As we show in the next subsection, we find that using the spectral radius is overly conservative, and in practice using the mean eigenvalue magnitude $\overline{|\lambda|}(J)=\frac{1}{N}\sum_{i=1}|\lambda_i|$ is sufficient to maintain high-quality solutions without unnecessarily high bandwidth.

\subsection{Optimization Performance}
In this section we explore how parallel execution configurations impact performance in unconstrained optimization problems. We focus on six $N=2000$ graph instances from the GSet~\cite{noauthor_index_nodate} MaxCut suite. We specifically focus on Erdős-Rényi (ER) uniform random graphs~\cite{erdos_evolution_1960} and Barabási-Albert (BA) scale-free graphs~\cite{albert_statistical_2002}. We test three graphs of each type ($\{\text{G27}, \text{G28}, \text{G29}\}$ and $\{\text{G39}, \text{G40}, \text{G41}\}$ respectively), each with bimodal $\{+1,-1\}$ couplings. We omit testing purely antiferromagnetic problems due to the unique spectral behavior explored on smaller instances in previous sections. 

\begin{figure}
    \centering
    \includegraphics[width=0.9\linewidth]{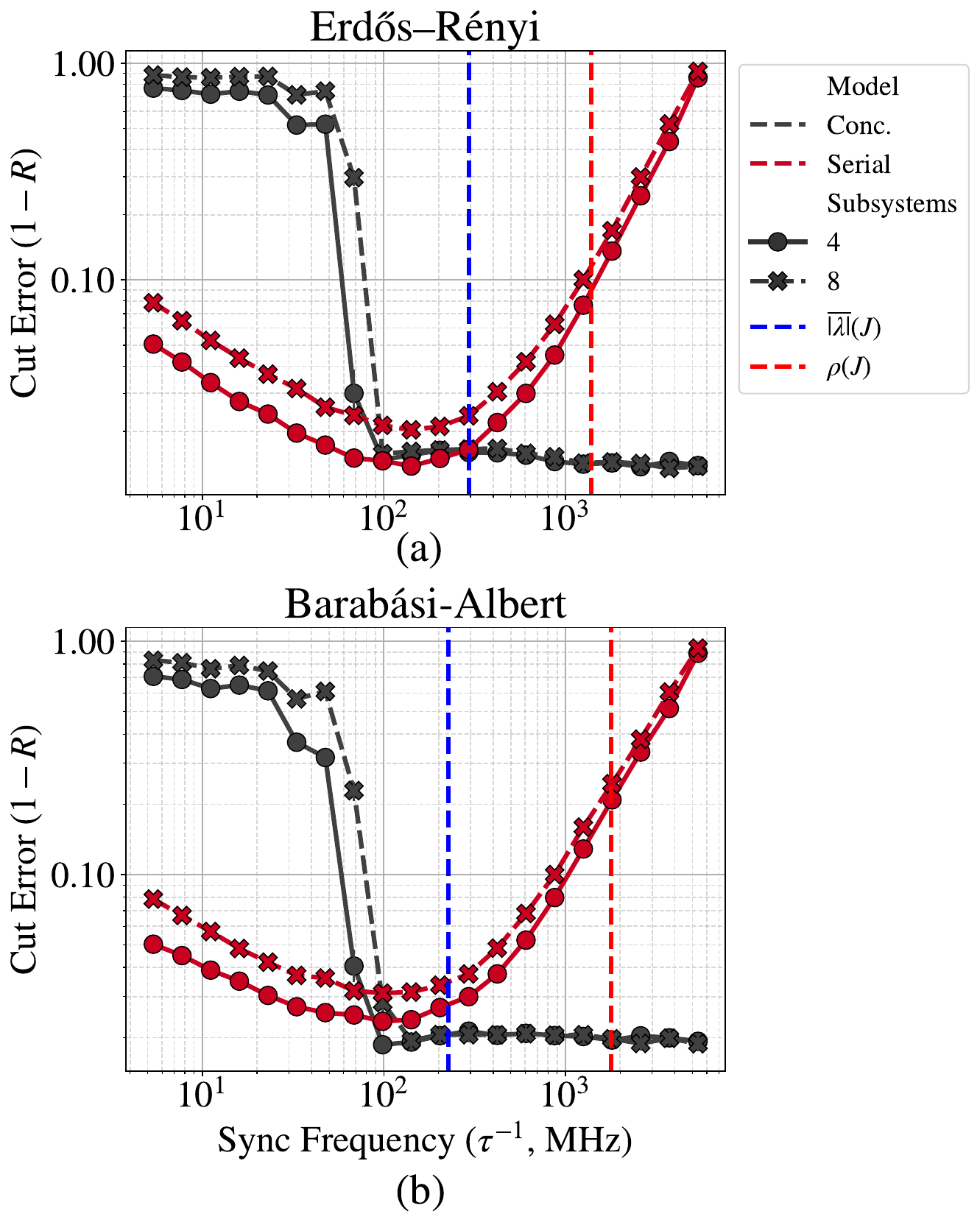}
    \caption{The cut error $1-R$ versus the synchronization frequency for 6 $N=2000$ (a) Erdős-Rényi (ER) and (b) Barabási-Albert (BA) graphs from the GSet~\cite{noauthor_index_nodate} suite.}
    \label{fig:er_ba}
\end{figure}

Fig.~\ref{fig:er_ba} shows the average solution quality versus synchronization frequency for several graphs from the GSet suite for serial and concurrent execution. We simulate each solver for 20 $\mu s$, with each datapoint representing 40 trials. The y-axis shows the ``Cut Error'' (lower=better) $1-\frac{Cut}{BKS}\triangleq 1-R$, where BKS is the best known solution for a given problem and $R$ is the ``approximation ratio''. We indicate the frequencies selected by the spectral radius $\rho(J)$ and average eigenvalue magnitude $\overline{|\lambda|}(J)$ with vertical lines.

For low frequencies ($<$100 MHz), both serial and concurrent models perform poorly (high cut errors). However, for serial mode, the cut error appears to be a smooth function of frequency, while the concurrent mode exhibits the sharp discontinuity seen in Fig.~\ref{fig:sk_energy_error}. As frequency increases, both models see solution quality gains up to the ``crossover point'' at approximately 100-200 MHz. Serial execution achieves higher quality solutions at the crossover point for $B=4$ on the ER graphs, while concurrent execution achieves lower average error rates for $B=8$ on both ER and BA as well as on $B=4$ BA graphs. 

Recall that the purpose of the  $\rho(J)$ and $\overline{|\lambda|}(J)$ heuristics is to provide an estimate of the high-frequency, ``high solution quality'' region, observed to be $>$100 MHz. Setting $\tau=\rho(J)$ yields a conservative estimate of 1400 MHz and 1800 MHz for ER and BA graphs respectively, $14-18\times$ greater than the actual cutoff value. In contrast, choosing $\tau=\overline{|\lambda|}(J)$ leads to significantly tighter estimates of 247 and 165 MHz for ER and BA graphs respectively. Using the more liberal heuristic therefore provides over $5\times$ and up to $11\times$ reductions in communication volume for the graphs tested while remaining within the region of high solution quality.

In the high frequency region ($>$100 MHz), the concurrent model solution quality is approximately constant, remaining at $\sim $0.015, while the serial mode error rapidly increases. The loss of solution quality is due to 1-bit quantization. After evolving for a $\tau$ second epoch, each serial subsystem state is clamped to $\pm 1$. Sufficiently small $\tau$ (high frequency) impede forward progress from the subsolver, as the system is not allowed to substantially change state before being binarized. Having higher precision spin state representations will mitigate this problem.

The data shown in Fig.~\ref{fig:er_ba} represent \emph{mean} energies, whereas optimization tasks are typically focused on finding (some approximation to) the optimal solution. Supposing we want to achieve some proportion of the best known solution (say 98\%), it is common practice to express performance in terms of ``metric to target'' (MTT)
\[\text{MTT}=\langle M\rangle\frac{\log(0.01)}{\log(1-p_{Success})}.\]
$MTT$ measures the estimated ``metric'' needed to achieve a target solution quality with 99\% probability. ``Metric'' is typically time, however in this work we also include energy as a principle cost. 

\begin{figure}[hbt]
    \centering
    \includegraphics[width=\linewidth]{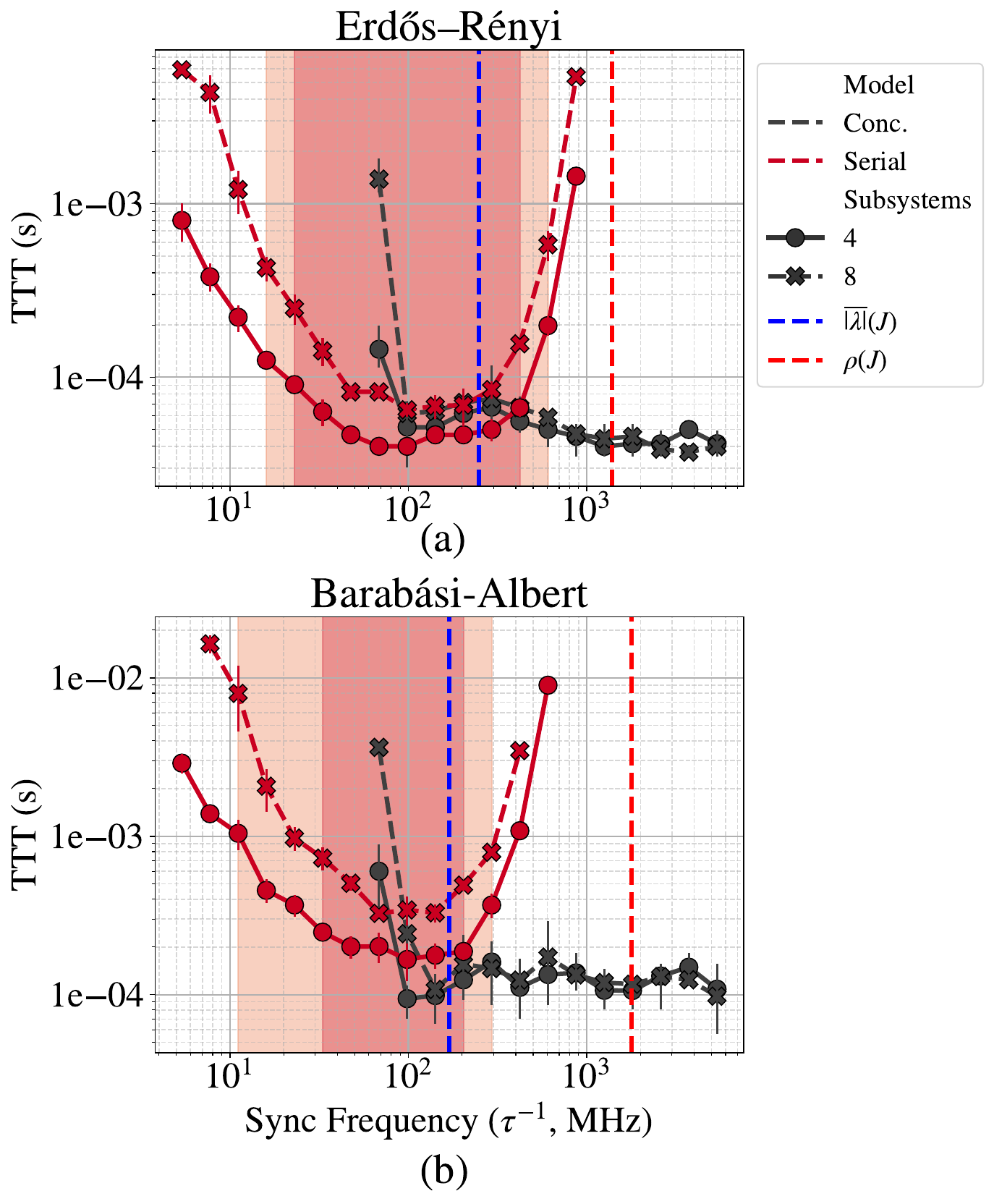}
    \caption{Mean time-to-target for the serial and concurrent models with increasing frequency (decreasing epoch $\tau$) on $N=2000$ GSet Barabási-Albert (a) and Erdős-Rényi (b) graphs. We also show the frequencies selected with the spectral heuristics, comparing the mean eigenvalue magnitude $\overline{|\lambda|}(J)$ (blue) and the spectral radius $\rho(J)$ (red). Shaded regions denote the $\tau$ values where the Serial model reached the solution quality target in $20\mu s$. Lighter regions denote $B=4$ and darker regions denote $B=8$.}
    \label{fig:er_ba_time}
\end{figure}

\begin{figure*}[bt]
    \centering
    \includegraphics[width=\linewidth]{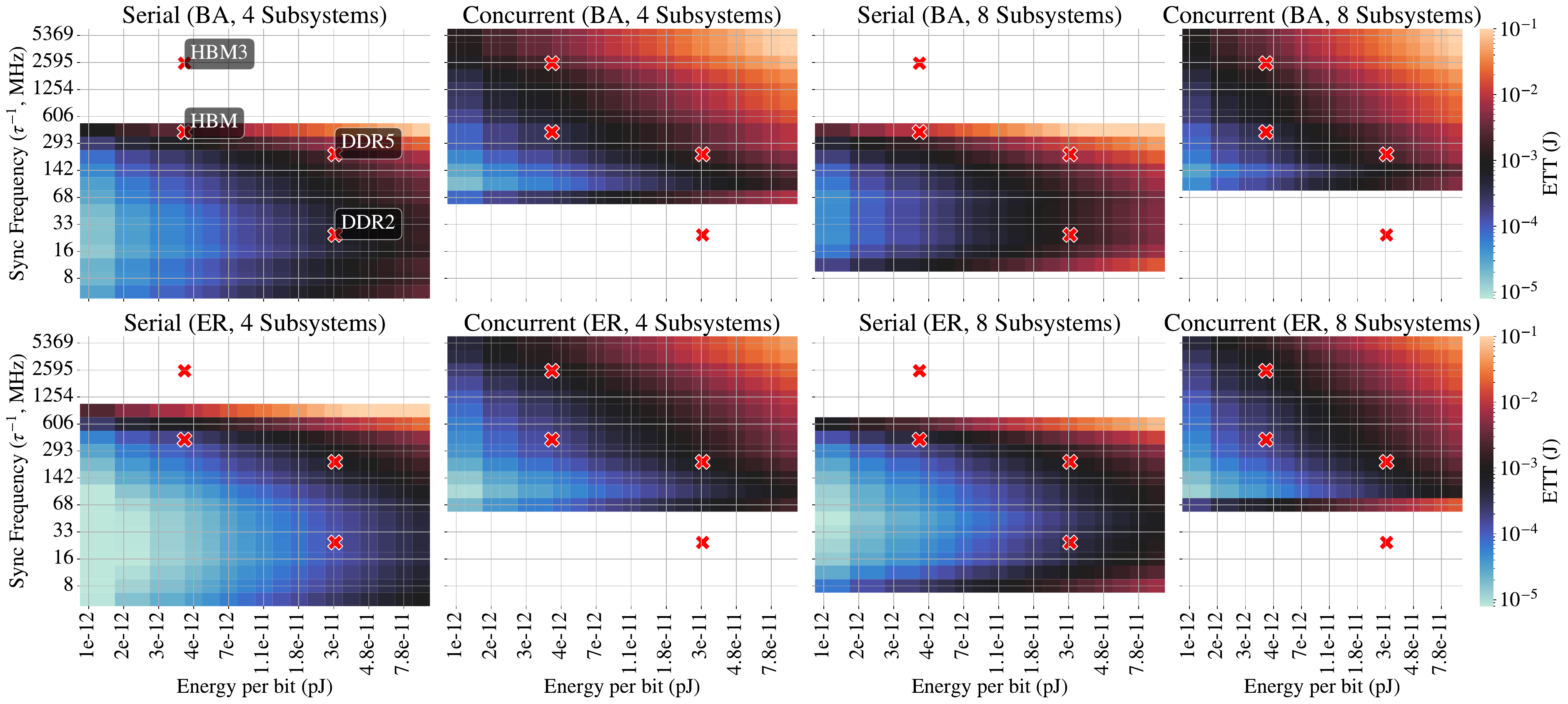}
    \caption{Mean estimated energy-to-target (ETT) versus synchronization frequency $\tau^{-1}$ and the energy cost per bit for serial and concurrent models on $N=2000$ GSet Barabási-Albert (top) and Erdős-Rényi (bottom) graphs. Markers indicating the maximum achievable frequency with modern memory interfaces are shown as red markers. Energy consumption is assumed to be approximately 34 and 4 pJ/bit for DRAM~\cite{giridhar_exploring_2013} and HBM~\cite{oconnor_fine-grained_2017} respectively. HBM/HBM3 bandwidths are assumed to be 1/5.6 Tbps~\cite{kim_present_2024} and DDR2/DDR5 bandwidths are 62/546 Gbps~\cite{noauthor_ram_nodate,noauthor_ddr5_nodate}. }
    \label{fig:er_ba_ene}
\end{figure*}

We set a solution quality target as 98\% of the best known solution (BKS), taken from Ref.~\cite{ma_multiple_2017}. Therefore the reported TTT/ETT is the time/energy required to reach 98\% of the BKS. Fig.~\ref{fig:er_ba_time} shows the TTT for serial and concurrent modes as a function of the update frequency. Missing values indicate that the target quality was not reached in 40 trials. Concurrent mode simulations lasted 20$\mu s$, however initial serial operation testing at $20 \mu s$ only reached the solution quality target for a narrow frequency band (shown as light and dark shaded bands for $B=4$ and $B=8$ respectively). As such, we also simulated the serial system for 40, 60, 80, and 100 $\mu s$ and took the lowest TTT figure for comparison. This methodology follows from the differences between each approach. In serial operation, $20\mu s$ of \emph{total} annealing time with 8 blocks results in only $2.5\mu s$ of annealing time per block. Therefore, increasing the annealing time follows from the necessary increase in latency required from adopting serial mode execution.

As in the mean solution quality results, serial execution outperformed concurrent in ER graphs, nearly matching high frequency concurrent TTT with update frequencies $\leq 200$ MHz. However, in BA graphs concurrent execution has over a $2\times$ TTT advantage for frequencies over 100 MHz. Increasing the subsystem count to $b=8$ leads to $\sim 10\times$ lower TTT for concurrent systems compared with serial execution. 

We also observe that in both mean cut error and TTT, selecting $\tau^{-1}=\rho(J)$ is overly conservative, requiring over $5\times$ higher frequency compared to $\tau^{-1}=\overline{|\lambda|}(J)$. The latter was within a factor of 2 of the optimal TTT point for both ER and BA graphs, leading to a superior frequency/TTT tradeoff. While the increased frequency requirement does not affect the latency, it has a marked effect on energy costs.

As we use 1 bit representations for $X_{0}$, we can also quantify the energy cost stemming from communication. Fig.~\ref{fig:er_ba_ene} shows the estimated energy-to-target (ETT) resulting from synchronization updates with increasing sync frequency $\tau^{-1}$ and energy per bit $E_{Bit}$. To put the required bandwidths/energies into perspective, we annotate the plot with memory interface techologies placed according to their peak throughput and estimated energy per bit (the latter taken from Ref~\cite{mccrabb_acre_2023}). 

We compute energy $E$ as
\begin{equation}
    E=2000E_{Bit}\left\lfloor\frac{t_{Anneal}}{\tau}\right\rfloor
\end{equation}
where $\left\lfloor\frac{t_{Anneal}}{\tau}\right\rfloor$ gives the number of synchronizations and $E_{Bit}$ is the energy per bit. We consider expended solver energy to be negligible compared to communication overhead, and moreover the expected accelerator energy would be similar for concurrent and serial architectures in any case.

Fig.~\ref{fig:er_ba_ene} illustrates the hidden cost of running high-frequency updates. The left hand subplots show serial/concurrent execution with 4 subsystems, while the RHS subplots show serial/concurrent execution with 8 subsystems. We plot ETT values for a range of $E_{Bit}$ values (x-axis), with markers indicating the (maximum $\tau^{-1}$, $E_{Bit}$) pairs for existing memory technologies. As with TTT, missing values indicate that the target solution quality was not reached in 40 trials.

While the time-to-target stays relatively constant for concurrent execution past 100 MHz, the energy cost increases linearly with frequency from communication overhead. Accordingly, $\tau=\rho(J)$ incurs a $10\times$ higher energy cost compared to $\tau=\overline{|\lambda|}(J)$. Below the 100 MHz threshold, the concurrent ETT abruptly diverges, while the serial ETT demonstrates a gradual upward trend. 

From the contours shown, optimal concurrent ETT is attained by minimizing the operating frequency without going below the stability threshold ($\sim$100 MHz). In contrast, optimal serial ETT occurs at lower frequencies ($\sim 50$ MHz), with a wider base of near-optimal frequencies. ETT therefore drastically differs from TTT, where concurrent operation has a distinct advantage (particularly in the 8 block case). The two approaches achieve similar ETT in the mid-frequency range ($\sim 70-500$ MHz) while serial operation extends the energy efficiency region into the low-frequency ($<50$ MHz) range. Concurrent operation is capable of recovering high-quality solutions for $>500$ MHz, however the marginal decrease in TTT is overwhelmed by the increasing energy cost of communication. The opposite is true of serial execution, where the increase in TTT at low frequencies is (somewhat) compensated by decreased communication costs, resulting in more graceful ETT degradation.

For efficient operation, Fig.~\ref{fig:er_ba_ene} implies that concurrent Ising machine architectures require an optimized synchronization schedule to achieve (near) optimal ETT; dependent on the target application and the power requirements of the application. Serial operation is observed to attain minimal ETT across a wider range of available bandwidths, allowing for more flexibility in parameter selection.

\begin{figure}[H]
    \centering
    \includegraphics[width=\linewidth]{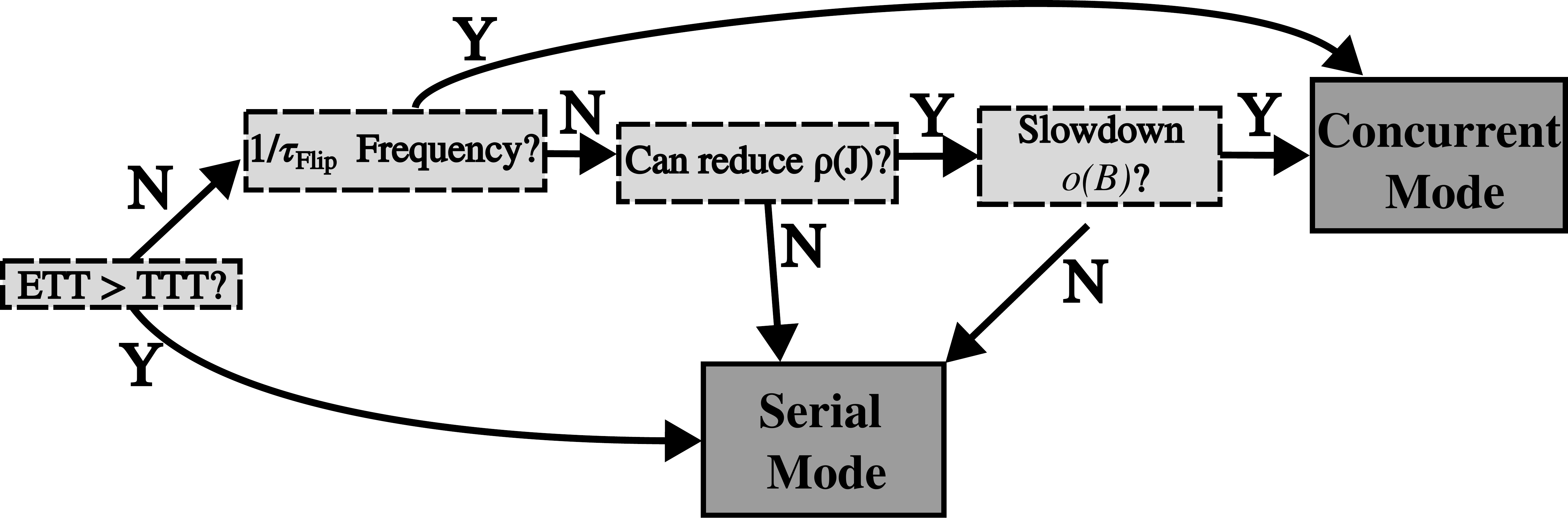}
    \caption{Parallel Ising machine operating mode decision tree. Note that most cases prefer serial operation to concurrent, however latency-critical applications with sufficiently high-available bandwidth can derive advantage from concurrent mode.}
    \label{fig:decision_tree}
\end{figure}

As our numerical and theoretical findings demonstrate, the gap between serial and concurrent mode operation is highly dependent on the synchronization epoch $\tau$ and the spectrum of $J$. Moreover, comparing Figs.~\ref{fig:er_ba_time} and~\ref{fig:er_ba_ene} suggests that comparisons between concurrent and serial are sensitive to the choice of metric (time vs. energy). Using the results gathered so far, we propose a simple decision tree to determine the operating mode most suitable to a given system/problem, shown in Fig.~\ref{fig:decision_tree}.

As Fig.~\ref{fig:er_ba_ene} shows, concurrent and serial operation achieve roughly the same ETT performance across energy-per-bit values, with energy efficiency peaking the $30-70$ MHz range. Given the additional theoretical and practical complications incurred by concurrent operation, Fig.~\ref{fig:decision_tree} shows that serial is preferable to concurrent operation in all scenarios where ETT is higher priority than TTT.

If TTT is preferred, then the recommendation depends on system parameters and problem structure. Eqn.~\ref{eqn:tau_flip} provides an empirically conservative heuristic for a maximal admissible synchronization period $\tau_{Flip}$ for effective concurrent operation based on the spectral radius $\rho(J)$ (or a more liberal substitute, such as the average eigenvalue magnitude). In our system model, $\rho(J)$ depends on both the problem structure (the problem couplings $\tilde{J}$) and on the system $R\cdot C$ time constant, related by $J=\frac{1}{R\cdot C}\tilde{J}$, therefore $\rho(J)=\frac{1}{R\cdot C}\tilde{J}$. 

If the proposed system cannot natively achieve $1/\tau_{Flip}$ synchronization frequencies, then the user can attempt to reduce $\rho(J)$, either by increasing the system $R\cdot C$ constant or by altering the problem formulation. For instance, lowering the penalty parameters in constrained Hamiltonian encodings lowers the spectral radius of the coupling matrix, though this must be considered in tandem with maintaining high feasibility rates. If the user is unable to alter system time constants or sufficiently change the problem formulation, then serial mode is preferable.

Even if the system time constant can be changed from $R\cdot C$ to $R'\cdot C'>R\cdot C$, doing so will likely incur a net slowdown in system performance. If that slowdown $\frac{R'\cdot C'}{R\cdot C}=o(B)$ (where $B$ is again the number of subsystems), then the performance edge of concurrent mode over serial will likely be maintained. However, if $\frac{R'\cdot C'}{R\cdot C} > B$ it is highly probable that the advantage of concurrent mode over serial has been lost, making serial again preferable by virtue of system simplicity and theoretical grounding. 

To summarize, if the user prioritizes latency over energy efficiency \emph{and all other system parameters are amenable}, then concurrent mode should be used. \textit{In all other cases}, serial operation is preferable.

\section{Conclusion}\label{sec:conclusion}
In this work we provide theoretical and numerical analyses of parallel analog Ising machine architectures. Noting the lack of existing results for concurrent execution models, we provide lower bounds on asymptotic KL-divergence  as well as proving sufficient conditions for per-iteration contraction in 1-Wasserstein distance. 

Numerical experiments on the linear IM model support our theoretical results, demonstrating that problem smoothness and synchronization frequency are primary determinants of parallel solver performance. We further characterize the energy and time trade-offs between serial and concurrent execution models, noting that concurrent architectures require optimized synchronization tuning to optimize energy efficiency while serial execution allows for energy-efficient operation across a range of update frequencies. Finally, we propose practical heuristics for users to decide between serial and concurrent execution, dependent on user preferences, problem structure, system dynamics, and bandwidth constraints.

\section*{Data Availability Statement}
The data that support the findings of this article are openly available~\cite{burns_mxburns2022parallel-ising-analysis_2025}.

\section*{Acknowledgments}
This work was supported in part by NSF under Awards No. 2231036 and No. 2233378; and by
DARPA under contract No. FA8650-23-C-7312. 
%\tableofcontents
\appendix
\begin{widetext}
\section{Proof of Proposition~\ref{prop:kl}}
\label{appdx:kl_proof}
\begin{proof}
Recall that for two probability measures $\mu$ and $\nu$ over $\region\subseteq\R^d$ with $\mu$ absolutely continuous w.r.t. $\nu$ the Radon-Nikodym derivative $\frac{d\mu}{d\nu}$ is an integrable function which satisfies
\begin{equation}
    \mu(A)=\int_A \nu(x)\frac{d\mu}{d\nu}(x)dx
\end{equation}
for any set $A\subseteq \region$. 
Let $\mu_t$, $\nu_t$ be defined as in the previous section. Since we are concerned with the behavior of continuous potentials over a bounded domain, we can assume the system gradients $\nabla U_{Int}$, $\nabla U_{Ext}$ are Lipschitz with some constant. Therefore we can express the Radon-Nikodym derivative using the multidimensional Girsanov formula (See Eqn.~7.138 of~\cite{liptser_statistics_2001})
\begin{equation}
\begin{split}
    \frac{d\mu_t}{d\nu_t}(x)=\exp[\frac{\beta}{2}\int_0^t \innerprod{\nabla U_{Int}(X_s)+\nabla U_{Ext}(X_0)-\nabla U(X_s)}{\nabla U(X_s)ds + \sqrt{2\beta^{-1}}dW_s}\\
    -\frac{\beta}{4}\int_0^t \innerprod{\nabla U(X_s)-\nabla U_{Int}(X_s)-\nabla U_{Ext}(X_0)}{\nabla U(X_s)+\nabla U_{Int}(X_s)+\nabla U_{Ext}(X_0)}ds]
\end{split}
\end{equation}
Disregarding the $dW_s$ term (since it will be disregarded in later expectation calculations), we have:
\begin{equation}
\begin{split}
    \frac{d\mu}{d\nu}(x)=\exp[\frac{\beta}{2}\int_0^t \innerprod{\nabla U_{Int}(X_s)+\nabla U_{Ext}(X_0)-\nabla U(X_s)}{\nabla U(X_s)}ds\\
    -\frac{\beta}{4}\int_0^t \innerprod{\nabla U_{Ext}(X_s)-\nabla U_{Ext}(X_0)}{2\nabla U(X_s)-\nabla U_{Ext}(X_s)+\nabla U_{Ext}(X_0)}ds]\\
    =\exp[
    -\frac{\beta}{4}\int_0^t \innerprod{\nabla U_{Ext}(X_s)-\nabla U_{Ext}(X_0)}{-\nabla U_{Ext}(X_s)+\nabla U_{Ext}(X_0)}ds]\\
    =\exp[
    \frac{\beta}{4}\int_0^t \norm{\nabla U_{Ext}(X_s)-\nabla U_{Ext}(X_0)}^2ds]
    =\exp[
    \frac{t\beta}{4}\langle \| \delta\nabla U\|^2\rangle_t]
\end{split}
\end{equation}
Then the KL-Divergence is given by
\begin{equation}
\begin{split}
    \kl{\mu_t}{\nu_t}=\int\mu_t(X) \frac{t\beta}{4}\langle \| \delta\nabla U\|^2\rangle_tdX\\
    =\mathbb{E}_{\mu_t} [\frac{t\beta}{4}\langle\| \delta\nabla U\|^2\rangle_t].
\end{split}
\end{equation}
    
\end{proof}
\section{Proof of Propositions~\ref{prop:w1_1} and~\ref{prop:w1_2} }
\label{appdx:w1_proof}
\subsection{Proof of Proposition~\ref{prop:w1_1}}
\begin{proof}
For simplicity we let $X_0=X^m$ be the initial state and consider the evolution between synchronization steps from time $t\in[0,\tau]$. The 1-Wasserstein distance between $\mu_t$ and $\nu_t$ is
\begin{equation}
\begin{split}
    W_1(\mu_t,\nu_t)&=\inf_{\gamma\in\mathcal{C}(\mu_t,\nu_t)}\mathbb{E}_\gamma\left[\onenorm{X_t-Y_t}\right]\\
    &=\inf_{\gamma\in\mathcal{C}(\mu_t,\nu_t)}\mathbb{E}_\gamma\left[\left\lVert\int_0^t \nabla U_{Int}(X_s)+\nabla U_{Ext}(X_0)- \nabla U_{Int}(Y_s)-\nabla U_{Ext}(Y_s)ds\right\rVert_1\right]\\
\end{split}
\end{equation}
which we can rearrange to form
\begin{equation}
    W_1(\mu_t,\nu_t)=\inf_{\gamma\in\mathcal{C}(\mu_t,\nu_t)}\mathbb{E}_\gamma\left[\left\lVert\int_0^t \nabla U(X_s)-\nabla U(Y_s)+\nabla U_{Ext}(X_0)-\nabla U_{Ext}(X_s)ds\right\rVert_1\right].
\end{equation}
By the optimality of $\gamma$, we can upper bound $W_1(\mu_t,\nu_t)$ by the transport distance using the product measure $\nu\times\mu$
\begin{equation}
    W_1(\mu_t,\nu_t)\leq\mathbb{E}_{\nu\times\mu}\left[\left\lVert\int_0^t \nabla U(X_s)-\nabla U(Y_s)+\nabla U_{Ext}(X_0)-\nabla U_{Ext}(X_s)ds\right\rVert_1\right].
\end{equation}
Using the triangle inequality and integrating each term and noting that $X_0=Y_0$, we obtain 
\begin{equation}
\begin{split}
    W_1(\mu_t,\nu_t)\leq&\mathbb{E}_{\nu\times\mu}\left\lVert\int_0^t \nabla U(X_s)-\nabla U(Y_s)dt\right\rVert_1+\mathbb{E}_{\nu\times\mu}\left\lVert\int_0^t\nabla U_{Ext}(X_0)-\nabla U_{Ext}(X_s)dt\right\rVert_1\\
    =&\mathbb{E}_{\nu\times\mu}\lVert\nabla U(Y_t)-\nabla U(X_t) \rVert_1+\mathbb{E}_{\mu}\lVert\nabla U_{Ext}(X_0)-\nabla U_{Ext}(X_t) \rVert_1\\
\end{split}
\end{equation}

Then we have an upper bound on $W_1(\mu, \nu)$ in terms of the iterates $X,Y$. Applying the triangle inequality, we obtain
\begin{equation}
\begin{split}
    W_1(\mu_t, \pi)\leq& W_1(\nu_t, \pi) + W_1(\nu_t,\mu_t)\\
                   \leq& \underbrace{C(t)W_1(\nu_0,\pi) + \mathbb{E}_{\nu\times\mu}\lVert\nabla U(Y_t)-\nabla U(X_t) \rVert_1+\mathbb{E}_{\mu}\lVert\nabla U_{Ext}(X_0)-\nabla U_{Ext}(X_t) \rVert_1}_{\text{(a)}}
\end{split}
\end{equation} 
If we have (a)$\;< W_1(\mu_0,\pi)$, then we have guaranteed contraction. With a bit of algebra, we obtain
\begin{equation}\label{eqn:appdx_w1_bound_tighter}
    \frac{\mathbb{E}_{\nu\times\mu}\lVert\nabla U(Y_t)-\nabla U(X_t) \rVert_1+\mathbb{E}_{\mu}\lVert\nabla U_{Ext}(X_0)-\nabla U_{Ext}(X_t) \rVert_1}{1-C(t)}<W_1(\nu_0,\pi)
\end{equation}
as a sufficient condition for contraction.
\end{proof}

\subsection{Proof of Proposition~\ref{prop:w1_2}}
We use Equation~\eqref{eqn:appdx_w1_bound_tighter} in numerical experiments, where we can empirically estimate $\nu_t$ and $\mu_t$. However, the behavior of the ideal process $Y_t$ is unknown in practice, making~\eqref{eqn:appdx_w1_bound_tighter} interesting, but not particularly useful. 

Assuming that the target potential is $L$-smooth, we provide a stronger condition which depends only on device characteristics, problem structure, and the concurrent process $X_t$. If we assume that the expected distance $\mathbb{E}_\gamma\onenorm{X_s-Y_s}$ is increasing with time and that $tL < 1$, we have
\begin{equation}
\begin{split}
    \frac{\mathbb{E}_{\mu_t}\onenorm{\nabla U_{Ext}(X_0)-\nabla U_{Ext}(X_t)}}{1-tL}<W_1(\mu_0,\nu_t).
\end{split}
\end{equation}

\begin{proof}
Applying the triangle inequality, we obtain
\begin{equation}
\begin{split}
    W_1(\mu_t,\nu_t) \leq\inf_{\gamma\in\mathcal{C}(\mu_t,\nu_t)}\mathbb{E}_\gamma\left[\onenorm{\int_0^t \nabla U(X_s)-\nabla U(Y_s)}ds+\onenorm{\int_0^t\nabla U_{Ext}(X_0)-\nabla U_{Ext}(X_s)ds}\right].
    \end{split}
    \end{equation}
Using Jensen's inequality and the smoothness of $U$, we obtain
    \begin{equation}
        \begin{split}
    W_1(\mu_t,\nu_t)\leq\inf_{\gamma\in\mathcal{C}(\mu_t,\nu_t)}\mathbb{E}_\gamma\left[\int_0^t L\onenorm{X_s-Y_s}ds+\onenorm{\int_0^t\nabla U_{Ext}(X_0)-\nabla U_{Ext}(X_s)ds}\right]\\
\end{split}
\end{equation}
Assuming that $\E_\gamma\onenorm{X_s-Y_s}$ is non-decreasing (i.e., $\sup_{s\in [0,t]}\E_\gamma\onenorm{X_s-Y_s}=\E_\gamma\onenorm{X_t-Y_t}$), we can apply the upper bound
\begin{equation}
\begin{split}
    W_1(\mu_t,\nu_t)\leq&\inf_{\gamma\in\mathcal{C}(\mu_t,\nu_t)}t\mathbb{E}_\gamma\left[L\onenorm{X_t-Y_t}+\onenorm{\int_0^t\nabla U_{Ext}(X_0)-\nabla U_{Ext}(X_s)ds}\right]\\
    =&tLW_1(\mu_t,\nu_t)+\mathbb{E}_{\mu_t}\onenorm{\nabla U_{Ext}(X_0)-\nabla U_{Ext}(X_t)}.
\end{split}
\end{equation}

For $tL< 1$, we obtain
\begin{equation}
     W_1(\mu_t,\nu_t)\leq \frac{\mathbb{E}_{\mu_t}\onenorm{\nabla U_{Ext}(X_0)-\nabla U_{Ext}(X_t)}}{1-tL}
\end{equation}
\end{proof}
\end{widetext}
\section{$W_1$ Experimental Details}\label{sec:appdx_experimental}
Here we provide experimental details for the numerical results in Section~\ref{sec:convergence}. Simulation code was written in Julia and run on an i9-13900k CPU with 64 GB of RAM. 

For a target problem, we randomly initialize $N=3000$ problem states from a uniform random distribution over $\{-1,1\}$. Each state is a different initialization of $X_0$, allowing for statistical averaging.

We then evolve two synchronously coupled processes, $X_t$ and $Y_t$, for $t\in [0, \tau]$ using an Euler-Maruyama discretization of~\eqref{eqn:sde_linear} with a step size of $dt=10^{-12}\text{ s}$. We use $R=\SI{310}{k\Omega}$ and $C=\SI{50}{fF}$ for device parameters, leading to an effective time constant of $\SI{15.5}{ns}$. After $\tau$ seconds, we quantize the states to the nearest hypercube vertex using the $\texttt{sign}$ function. At $\beta=10$, the continuous processes strongly concentrated around the hypercube vertices, making the quantized state a reasonable approximation of the continuous state.

The quantized states are used to compute an empirical distribution over the $2^{12}=4096$ vertices of the hypercube, forming a distribution $\mu\in\R^{4096}$. We also compute the true distribution $\pi$ by fully enumerating the spin glass states. Using the package PythonOT.jl (the Julia interface for PythonOT~\cite{flamaryPOTPythonOptimal2021b}), we compute the $W_1$ distance between the distributions with the \texttt{emd2} function. The distance matrix was computed using pairwise Hamming distances between states.

Our target graph was a 12-spin ferromagnetic system, arranged in a two-dimensional, periodic $4\times 3$ lattice. Despite our broader focus on non-convex problems, we selected a ferromagnetic instance to avoid the high estimation variance resulting from solver samples being stuck in local minima. 
%\pagebreak
\bibliography{refs1}% Produces the bibliography via BibTeX.

@inproceedings{sundara_raman_sachi_2024,
	title = {{SACHI}: {A} {Stationarity}-{Aware}, {All}-{Digital}, {Near}-{Memory}, {Ising} {Architecture}},
	shorttitle = {{SACHI}},
	url = {https://ieeexplore.ieee.org/abstract/document/10476402},
	doi = {10.1109/HPCA57654.2024.00061},
	urldate = {2024-07-27},
	booktitle = {2024 {IEEE} {International} {Symposium} on {High}-{Performance} {Computer} {Architecture} ({HPCA})},
	author = {Sundara Raman, Siddhartha Raman and John, Lizy K. and Kulkarni, Jaydeep P.},
	month = mar,
	year = {2024},
	note = {ISSN: 2378-203X},
	pages = {719--731},
}

@misc{mcgeoch_d-wave_2020,
	title = {D-{Wave} {Hybrid} {Solver} {Service} + {Advantage}: {Technology} {Update}},
	author = {McGeoch, Catherine and Farre, Pau and Bernoudy, William},
	month = sep,
	year = {2020},
        howpublished={Available: [Online] \url{https://www.dwavequantum.com/media/m2xbmlhs/14-1048a-a_d-wave_hybrid_solver_service_plus_advantage_technology_update.pdf}},
}

@inproceedings{vempala_rapid_2019,
	title = {Rapid {Convergence} of the {Unadjusted} {Langevin} {Algorithm}: {Isoperimetry} {Suffices}},
	volume = {32},
	shorttitle = {Rapid {Convergence} of the {Unadjusted} {Langevin} {Algorithm}},
	url = {https://proceedings.neurips.cc/paper/2019/hash/65a99bb7a3115fdede20da98b08a370f-Abstract.html},
	urldate = {2024-02-27},
	booktitle = {Advances in {Neural} {Information} {Processing} {Systems}},
	publisher = {Curran Associates, Inc.},
	author = {Vempala, Santosh and Wibisono, Andre},
	year = {2019},
}

@article{durmus_efficient_2016,
  title = {Efficient {{Bayesian Computation}} by {{Proximal Markov Chain Monte Carlo}}: {{When Langevin Meets Moreau}}},
  shorttitle = {Efficient {{Bayesian Computation}} by {{Proximal Markov Chain Monte Carlo}}},
  author = {Durmus, Alain and Moulines, {\'E}ric and Pereyra, Marcelo},
  year = {2018},
  month = jan,
  journal = {SIAM Journal on Imaging Sciences},
  volume = {11},
  number = {1},
  pages = {473--506},
  publisher = {{Society for Industrial and Applied Mathematics}},
  doi = {10.1137/16M1108340},
  urldate = {2025-06-11}
}

@article{boyd_distributed_2011,
	title = {Distributed {Optimization} and {Statistical} {Learning} via the {Alternating} {Direction} {Method} of {Multipliers}},
	volume = {3},
	issn = {1935-8237, 1935-8245},
	url = {https://www.nowpublishers.com/article/Details/MAL-016},
	doi = {10.1561/2200000016},
	number = {1},
	urldate = {2024-04-27},
	journal = {Foundations and Trends® in Machine Learning},
	author = {Boyd, Stephen and Parikh, Neal and Chu, Eric and Peleato, Borja and Eckstein, Jonathan},
	month = jul,
	year = {2011},
	note = {Publisher: Now Publishers, Inc.},
	pages = {1--122},
}

@inproceedings{mccrabb_acre_2023,
	address = {Toronto ON Canada},
	title = {{ACRE}: {Accelerating} {Random} {Forests} for {Explainability}},
	isbn = {9798400703294},
	shorttitle = {{ACRE}},
	url = {https://dl.acm.org/doi/10.1145/3613424.3623788},
	doi = {10.1145/3613424.3623788},
	urldate = {2024-04-06},
	booktitle = {56th {Annual} {IEEE}/{ACM} {International} {Symposium} on {Microarchitecture}},
	publisher = {ACM},
	author = {McCrabb, Andrew and Ahmed, Aymen and Bertacco, Valeria},
	month = oct,
	year = {2023},
	pages = {1016--1028},
}

@misc{cheng_sharp_2020,
	title = {Sharp convergence rates for {Langevin} dynamics in the nonconvex setting},
	url = {http://arxiv.org/abs/1805.01648},
	urldate = {2023-12-06},
	publisher = {arXiv},
	author = {Cheng, Xiang and Chatterji, Niladri S. and Abbasi-Yadkori, Yasin and Bartlett, Peter L. and Jordan, Michael I.},
	month = jul,
	year = {2020},
	note = {arXiv:1805.01648 [cs, math, stat]},
}

@article{ma_multiple_2017,
	title = {A multiple search operator heuristic for the max-k-cut problem},
	volume = {248},
	issn = {1572-9338},
	url = {https://doi.org/10.1007/s10479-016-2234-0},
	doi = {10.1007/s10479-016-2234-0},
	number = {1},
	urldate = {2023-11-08},
	journal = {Annals of Operations Research},
	author = {Ma, Fuda and Hao, Jin-Kao},
	month = jan,
	year = {2017},
	pages = {365--403},
}

@misc{booth_partitioning_nodate,
	title = {Partitioning {Optimization} {Problems} for {Hybrid} {Classical}/{Quantum} {Execution}},
	author = {Booth, Michael and Reinhardt, Steven P and Roy, Aidan},
        howpublished={Available: [Online] \url{https://www.dwavequantum.com/media/jhlpvult/partitioning_qubos_for_quantum_acceleration-2.pdf}},
}

@article{raymond_hybrid_2023,
  title={Hybrid quantum annealing for larger-than-QPU lattice-structured problems},
  author={Raymond, Jack and Stevanovic, Radomir and Bernoudy, William and Boothby, Kelly and McGeoch, Catherine C and Berkley, Andrew J and Farr{\'e}, Pau and Pasvolsky, Joel and King, Andrew D},
  journal={ACM Transactions on Quantum Computing},
  volume={4},
  number={3},
  pages={1--30},
  year={2023},
  publisher={ACM New York, NY, USA}
}

@inproceedings{tan_hyqsat_2023,
	title = {{HyQSAT}: {A} {Hybrid} {Approach} for 3-{SAT} {Problems} by {Integrating} {Quantum} {Annealer} with {CDCL}},
	shorttitle = {{HyQSAT}},
	doi = {10.1109/HPCA56546.2023.10071022},
	booktitle = {2023 {IEEE} {International} {Symposium} on {High}-{Performance} {Computer} {Architecture} ({HPCA})},
	author = {Tan, Siwei and Yu, Mingqian and Python, Andre and Shang, Yongheng and Li, Tingting and Lu, Liqiang and Yin, Jianwei},
	month = feb,
	year = {2023},
	note = {ISSN: 2378-203X},
	pages = {731--744},
}

@inproceedings{afoakwa_brim_2021,
	title = {{BRIM}: {Bistable} {Resistively}-{Coupled} {Ising} {Machine}},
	shorttitle = {{BRIM}},
	doi = {10.1109/HPCA51647.2021.00068},
	booktitle = {2021 {IEEE} {International} {Symposium} on {High}-{Performance} {Computer} {Architecture} ({HPCA})},
	author = {Afoakwa, Richard and Zhang, Yiqiao and Vengalam, Uday Kumar Reddy and Ignjatovic, Zeljko and Huang, Michael},
	month = feb,
	year = {2021},
	note = {ISSN: 2378-203X},
	pages = {749--760},
}

@inproceedings{sharma_increasing_2022,
	address = {New York, NY, USA},
	series = {{ISCA} '22},
	title = {Increasing ising machine capacity with multi-chip architectures},
	isbn = {978-1-4503-8610-4},
	url = {https://dl.acm.org/doi/10.1145/3470496.3527414},
	doi = {10.1145/3470496.3527414},
	urldate = {2023-05-18},
	booktitle = {Proceedings of the 49th {Annual} {International} {Symposium} on {Computer} {Architecture}},
	publisher = {Association for Computing Machinery},
	author = {Sharma, Anshujit and Afoakwa, Richard and Ignjatovic, Zeljko and Huang, Michael},
	month = jun,
	year = {2022},
	pages = {508--521},
}

@article{chiang_diffusion_1987,
	title = {Diffusion for {Global} {Optimization} in $\mathbb{R}^n$},
	volume = {25},
	issn = {0363-0129},
	url = {https://epubs.siam.org/doi/10.1137/0325042},
	doi = {10.1137/0325042},
	number = {3},
	urldate = {2023-06-05},
	journal = {SIAM Journal on Control and Optimization},
	author = {Chiang, Tzuu-Shuh and Hwang, Chii-Ruey and Sheu, Shuenn Jyi},
	month = may,
	year = {1987},
	note = {Publisher: Society for Industrial and Applied Mathematics},
	pages = {737--753},
}

@article{wang_oim_2019,
	title = {{OIM}: {Oscillator}-based {Ising} {Machines} for {Solving} {Combinatorial} {Optimisation} {Problems}},
	volume = {abs/1903.07163},
	url = {http://arxiv.org/abs/1903.07163},
	journal = {CoRR},
	author = {Wang, Tianshi and Roychowdhury, Jaijeet},
	year = {2019},
	note = {\_eprint: 1903.07163},
}

@article{lucas_ising_2014,
	title = {Ising formulations of many {NP} problems},
	volume = {2},
	issn = {2296-424X},
	url = {https://www.frontiersin.org/articles/10.3389/fphy.2014.00005},
	urldate = {2023-04-22},
	journal = {Frontiers in Physics},
	author = {Lucas, Andrew},
	year = {2014},
}

@article{barahona_computational_1982,
	title = {On the computational complexity of {Ising} spin glass models},
	volume = {15},
	issn = {0305-4470},
	url = {https://dx.doi.org/10.1088/0305-4470/15/10/028},
	doi = {10.1088/0305-4470/15/10/028},
	number = {10},
	urldate = {2023-08-13},
	journal = {Journal of Physics A: Mathematical and General},
	author = {Barahona, F.},
	month = oct,
	year = {1982},
	pages = {3241},
}

@misc{li_sharp_2022,
	title = {A sharp uniform-in-time error estimate for {Stochastic} {Gradient} {Langevin} {Dynamics}},
	url = {http://arxiv.org/abs/2207.09304},
	doi = {10.48550/arXiv.2207.09304},
	urldate = {2023-09-05},
	publisher = {arXiv},
	author = {Li, Lei and Wang, Yuliang},
	month = oct,
	year = {2022},
	note = {arXiv:2207.09304 [cs, math, stat]},
}

@article{albert_statistical_2002,
	title = {Statistical mechanics of complex networks},
	volume = {74},
	url = {https://link.aps.org/doi/10.1103/RevModPhys.74.47},
	doi = {10.1103/RevModPhys.74.47},
	number = {1},
	urldate = {2023-09-05},
	journal = {Reviews of Modern Physics},
	author = {Albert, Réka and Barabási, Albert-László},
	month = jan,
	year = {2002},
	note = {Publisher: American Physical Society},
	pages = {47--97},
}

@inproceedings{istrailStatisticalMechanicsThreedimensionality2000,
  title = {Statistical Mechanics, Three-Dimensionality and {{NP-completeness}}: {{I}}. {{Universality}} of Intracatability for the Partition Function of the {{Ising}} Model across Non-Planar Surfaces (Extended Abstract)},
  shorttitle = {Statistical Mechanics, Three-Dimensionality and {{NP-completeness}}},
  booktitle = {Proceedings of the Thirty-Second Annual {{ACM}} Symposium on {{Theory}} of Computing},
  author = {Istrail, Sorin},
  year = {2000},
  month = may,
  series = {{{STOC}} '00},
  pages = {87--96},
  publisher = {Association for Computing Machinery},
  address = {New York, NY, USA},
  doi = {10.1145/335305.335316},
  urldate = {2025-05-26},
  isbn = {978-1-58113-184-0}
}

@article{zhang_review_2024,
	title = {A {Review} of {Ising} {Machines} {Implemented} in {Conventional} and {Emerging} {Technologies}},
	volume = {23},
	issn = {1941-0085},
	url = {https://ieeexplore.ieee.org/document/10670493},
	doi = {10.1109/TNANO.2024.3457533},
	urldate = {2024-11-25},
	journal = {IEEE Transactions on Nanotechnology},
	author = {Zhang, Tingting and Tao, Qichao and Liu, Bailiang and Grimaldi, Andrea and Raimondo, Eleonora and Jiménez, Manuel and Avedillo, María José and Nuñez, Juan and Linares-Barranco, Bernabé and Serrano-Gotarredona, Teresa and Finocchio, Giovanni and Han, Jie},
	year = {2024},
	note = {Conference Name: IEEE Transactions on Nanotechnology},
	pages = {704--717},
}

@article{albash_adiabatic_2018,
	title = {Adiabatic quantum computation},
	volume = {90},
	url = {https://link.aps.org/doi/10.1103/RevModPhys.90.015002},
	doi = {10.1103/RevModPhys.90.015002},
	number = {1},
	journal = {Rev. Mod. Phys.},
	author = {Albash, Tameem and Lidar, Daniel A.},
	month = jan,
	year = {2018},
	note = {Publisher: American Physical Society},
	pages = {015002},
}

@article{hauke_perspectives_2020,
	title = {Perspectives of quantum annealing: methods and implementations},
	volume = {83},
	issn = {0034-4885},
	shorttitle = {Perspectives of quantum annealing},
	url = {https://dx.doi.org/10.1088/1361-6633/ab85b8},
	doi = {10.1088/1361-6633/ab85b8},
	
	number = {5},
	urldate = {2023-06-23},
	journal = {Reports on Progress in Physics},
	author = {Hauke, Philipp and Katzgraber, Helmut G. and Lechner, Wolfgang and Nishimori, Hidetoshi and Oliver, William D.},
	month = may,
	year = {2020},
	note = {Publisher: IOP Publishing},
	pages = {054401},
}

@article{matsumoto_distance-based_2022-2,
	title = {Distance-based clustering using {QUBO} formulations},
	volume = {12},
	copyright = {2022 The Author(s)},
	issn = {2045-2322},
	url = {https://www.nature.com/articles/s41598-022-06559-z},
	doi = {10.1038/s41598-022-06559-z},
	
	number = {1},
	urldate = {2023-08-25},
	journal = {Scientific Reports},
	author = {Matsumoto, Nasa and Hamakawa, Yohei and Tatsumura, Kosuke and Kudo, Kazue},
	month = feb,
	year = {2022},
	note = {Number: 1
Publisher: Nature Publishing Group},
	pages = {2669},
}

@inproceedings{vengalam_supporting_2023,
	address = {New York, NY, USA},
	series = {{MICRO} '23},
	title = {{SUPPORTING} {ENERGY}-{BASED} {LEARNING} {WITH} {AN} {ISING} {MACHINE} {SUBSTRATE}: {A} {CASE} {STUDY} {ON} {RBM}},
	isbn = {9798400703294},
	shorttitle = {{SUPPORTING} {ENERGY}-{BASED} {LEARNING} {WITH} {AN} {ISING} {MACHINE} {SUBSTRATE}},
	url = {https://dl.acm.org/doi/10.1145/3613424.3614315},
	doi = {10.1145/3613424.3614315},
	urldate = {2024-02-15},
	booktitle = {Proceedings of the 56th {Annual} {IEEE}/{ACM} {International} {Symposium} on {Microarchitecture}},
	publisher = {Association for Computing Machinery},
	author = {Vengalam, Uday Kumar Reddy and Liu, Yongchao and Geng, Tong and Wu, Hui and Huang, Michael},
	month = dec,
	year = {2023},
	pages = {465--478},
}

@book{chewiStatisticalOptimalTransport2025,
  title = {Statistical {{Optimal Transport}}: {{{\'E}cole}} d'{{{\'E}t{\'e}}} de {{Probabilit{\'e}s}} de {{Saint-Flour XLIX}} -- 2019},
  shorttitle = {Statistical {{Optimal Transport}}},
  author = {Chewi, Sinho and {Niles-Weed}, Jonathan and Rigollet, Philippe},
  year = {2025},
  series = {Lecture {{Notes}} in {{Mathematics}}},
  volume = {2364},
  publisher = {Springer Nature Switzerland},
  address = {Cham},
  doi = {10.1007/978-3-031-85160-5},
  urldate = {2025-06-11},
  copyright = {https://www.springernature.com/gp/researchers/text-and-data-mining},
  isbn = {978-3-031-85159-9 978-3-031-85160-5},
  langid = {english}
}

@article{flamaryPOTPythonOptimal2021b,
  title = {{{POT}}: {{Python Optimal Transport}}},
  shorttitle = {{{POT}}},
  author = {Flamary, R{\'e}mi and Courty, Nicolas and Gramfort, Alexandre and Alaya, Mokhtar Z. and Boisbunon, Aur{\'e}lie and Chambon, Stanislas and Chapel, Laetitia and Corenflos, Adrien and Fatras, Kilian and Fournier, Nemo and Gautheron, L{\'e}o and Gayraud, Nathalie T. H. and Janati, Hicham and Rakotomamonjy, Alain and Redko, Ievgen and Rolet, Antoine and Schutz, Antony and Seguy, Vivien and Sutherland, Danica J. and Tavenard, Romain and Tong, Alexander and Vayer, Titouan},
  year = {2021},
  journal = {Journal of Machine Learning Research},
  volume = {22},
  number = {78},
  pages = {1--8},
  issn = {1533-7928},
  urldate = {2025-06-11}
}

@book{villaniTopicsOptimalTransportation2003,
  title = {Topics in {{Optimal Transportation}}},
  author = {Villani, Cedric},
  year = {2003},
  month = mar,
  edition = {UK ed. edition},
  publisher = {American Mathematical Society},
  address = {Providence, RI},
  isbn = {978-0-8218-3312-4},
  langid = {english}
}

@article{elmitwalli_utilizing_2024,
	title = {Utilizing {Multi}-{Body} {Interactions} in a {CMOS}-{Based} {Ising} {Machine} for {LDPC} {Decoding}},
	volume = {71},
	issn = {1558-0806},
	url = {https://ieeexplore.ieee.org/abstract/document/10285565},
	doi = {10.1109/TCSI.2023.3322325},
	number = {1},
	urldate = {2024-03-19},
	journal = {IEEE Transactions on Circuits and Systems I: Regular Papers},
	author = {Elmitwalli, Eslam and Ignjatovic, Zeljko and Kose, Selcuk},
	month = jan,
	year = {2024},
	note = {Conference Name: IEEE Transactions on Circuits and Systems I: Regular Papers},
	pages = {40--50},
}

@misc{burns_mxburns2022parallel-ising-analysis_2025,
	title = {{Parallel}-{Ising}-{Analysis}},
	url = {https://github.com/mxburns2022/Parallel-Ising-Analysis},
	howpublished = {\url{https://github.com/mxburns2022/Parallel-Ising-Analysis}},
	urldate = {2025-03-25},
	author = {Burns, Matthew X.},
	month = mar,
	year = {2025},
}

@article{mohseni_ising_2022,
	title = {Ising machines as hardware solvers of combinatorial optimization problems},
	volume = {4},
	issn = {2522-5820},
	url = {https://www.nature.com/articles/s42254-022-00440-8},
	doi = {10.1038/s42254-022-00440-8},
	
	number = {6},
	urldate = {2024-10-16},
	journal = {Nature Reviews Physics},
	author = {Mohseni, Naeimeh and McMahon, Peter L. and Byrnes, Tim},
	month = may,
	year = {2022},
	pages = {363--379},
}

@article{inagaki_coherent_2016,
	title = {A coherent {Ising} machine for 2000-node optimization problems},
	volume = {354},
	url = {https://www.science.org/doi/full/10.1126/science.aah4243},
	doi = {10.1126/science.aah4243},
	number = {6312},
	urldate = {2023-08-13},
	journal = {Science},
	author = {Inagaki, Takahiro and Haribara, Yoshitaka and Igarashi, Koji and Sonobe, Tomohiro and Tamate, Shuhei and Honjo, Toshimori and Marandi, Alireza and McMahon, Peter L. and Umeki, Takeshi and Enbutsu, Koji and Tadanaga, Osamu and Takenouchi, Hirokazu and Aihara, Kazuyuki and Kawarabayashi, Ken-ichi and Inoue, Kyo and Utsunomiya, Shoko and Takesue, Hiroki},
	month = nov,
	year = {2016},
	note = {Publisher: American Association for the Advancement of Science},
	pages = {603--606},
}

@article{jearl_parallel_2005,
	title = {Parallel tempering: {Theory}, applications, and new perspectives},
	volume = {7},
	shorttitle = {Parallel tempering},
	url = {https://pubs.rsc.org/en/content/articlelanding/2005/cp/b509983h},
	doi = {10.1039/B509983H},
	
	number = {23},
	urldate = {2023-05-18},
	journal = {Physical Chemistry Chemical Physics},
	author = {J. Earl, David and W. Deem, Michael},
	year = {2005},
	note = {Publisher: Royal Society of Chemistry},
	pages = {3910--3916},
}

@misc{reference_workflows_dwave,
	title = {Reference {Workflows} — {Ocean} {Documentation} 8.2.0 documentation},
	url = {https://docs.ocean.dwavesys.com/en/latest/docs_hybrid/reference/reference.html},
	urldate = {2025-03-19},
howpublished = {\url{https://docs.ocean.dwavesys.com/en/latest/docs_hybrid/reference/reference.html}}
}

@inproceedings{giridhar_exploring_2013,
	address = {Denver Colorado},
	title = {Exploring {DRAM} organizations for energy-efficient and resilient exascale memories},
	isbn = {978-1-4503-2378-9},
	url = {https://dl.acm.org/doi/10.1145/2503210.2503215},
	doi = {10.1145/2503210.2503215},
	
	urldate = {2025-03-19},
	booktitle = {Proceedings of the {International} {Conference} on {High} {Performance} {Computing}, {Networking}, {Storage} and {Analysis}},
	publisher = {ACM},
	author = {Giridhar, Bharan and Cieslak, Michael and Duggal, Deepankar and Dreslinski, Ronald and Chen, Hsing Min and Patti, Robert and Hold, Betina and Chakrabarti, Chaitali and Mudge, Trevor and Blaauw, David},
	month = nov,
	year = {2013},
	pages = {1--12},
}

@inproceedings{kim_present_2024,
	title = {Present and {Future}, {Challenges} of {High} {Bandwith} {Memory} ({HBM})},
	url = {https://ieeexplore.ieee.org/document/10536972/?arnumber=10536972},
	doi = {10.1109/IMW59701.2024.10536972},
	urldate = {2025-03-19},
	booktitle = {2024 {IEEE} {International} {Memory} {Workshop} ({IMW})},
	author = {Kim, Kwiwook and Park, Myeong-jae},
	month = may,
	year = {2024},
	note = {ISSN: 2573-7503},
	pages = {1--4},
}

@article{fangParallelTemperingSimulation2014,
  title = {Parallel Tempering Simulation of the Three-Dimensional {{Edwards}}--{{Anderson}} Model with Compact Asynchronous Multispin Coding on {{GPU}}},
  author = {Fang, Ye and Feng, Sheng and Tam, Ka-Ming and Yun, Zhifeng and Moreno, Juana and Ramanujam, J. and Jarrell, Mark},
  year = {2014},
  month = oct,
  journal = {Computer Physics Communications},
  volume = {185},
  number = {10},
  pages = {2467--2478},
  issn = {00104655},
  doi = {10.1016/j.cpc.2014.05.020},
  urldate = {2025-06-06},
  langid = {english}
}

@incollection{sokalMonteCarloMethods1997,
  title = {Monte {{Carlo Methods}} in {{Statistical Mechanics}}: {{Foundations}} and {{New Algorithms}}},
  shorttitle = {Monte {{Carlo Methods}} in {{Statistical Mechanics}}},
  booktitle = {Functional {{Integration}}: {{Basics}} and {{Applications}}},
  author = {Sokal, A.},
  editor = {{DeWitt-Morette}, Cecile and Cartier, Pierre and Folacci, Antoine},
  year = {1997},
  pages = {131--192},
  publisher = {Springer US},
  address = {Boston, MA},
  doi = {10.1007/978-1-4899-0319-8_6},
  urldate = {2025-06-04},
  isbn = {978-1-4899-0319-8},
  langid = {english}
}

@book{newmanMonteCarloMethods1999,
  title = {Monte {{Carlo Methods}} in {{Statistical Physics}}},
  author = {Newman, M. E. J. and Barkema, G. T.},
  year = {1999},
  month = apr,
  publisher = {Oxford University Press},
  address = {Oxford, New York},
  isbn = {978-0-19-851797-9}
}

@article{ogielskiDynamicsThreedimensionalIsing1985,
  title = {Dynamics of Three-Dimensional {{Ising}} Spin Glasses in Thermal Equilibrium},
  author = {Ogielski, Andrew T.},
  year = {1985},
  month = dec,
  journal = {Physical Review B},
  volume = {32},
  number = {11},
  pages = {7384--7398},
  publisher = {American Physical Society},
  doi = {10.1103/PhysRevB.32.7384},
  urldate = {2025-06-06}
}

@inproceedings{oconnor_fine-grained_2017,
	title = {Fine-{Grained} {DRAM}: {Energy}-{Efficient} {DRAM} for {Extreme} {Bandwidth} {Systems}},
	shorttitle = {Fine-{Grained} {DRAM}},
	url = {https://ieeexplore.ieee.org/document/8686544/?arnumber=8686544},
	urldate = {2025-03-19},
	booktitle = {2017 50th {Annual} {IEEE}/{ACM} {International} {Symposium} on {Microarchitecture} ({MICRO})},
	author = {O'Connor, Mike and Chatterjee, Niladrish and Lee, Donghyuk and Wilson, John and Agrawal, Aditya and Keckler, Stephen W. and Dally, William J.},
	month = oct,
	year = {2017},
	note = {ISSN: 2379-3155},
	pages = {41--54},
}

@misc{noauthor_ddr5_nodate,
	title = {{DDR5} {RAM}: {Everything} {You} {Need} to {Know}},
	shorttitle = {{DDR5} {RAM}},
	url = {https://www.crucial.com/articles/about-memory/everything-about-ddr5-ram},
	howpublished={[Online] Available: \url{https://www.crucial.com/articles/about-memory/everything-about-ddr5-ram}},
        publisher={Crucial},
	urldate = {2025-03-19},
        year={2025},
	journal = {Crucial},
}

@misc{noauthor_ram_nodate,
	title = {{RAM} {Memory} {Speeds} \& {Compatibility}},
	url = {https://www.crucial.com/support/memory-speeds-compatability},
	howpublished={[Online] Available: \url{https://www.crucial.com/support/memory-speeds-compatability}},
	urldate = {2025-03-19},
	journal = {Crucial},
}

@misc{noauthor_index_nodate,
	title = {GSet Benchmark},
	url = {https://web.stanford.edu/~yyye/yyye/Gset/},
        howpublished = {[Online] Available: \url{https://web.stanford.edu/~yyye/yyye/Gset/}},
        author={Y. Ye},
	urldate = {2023-10-17},
}

@inproceedings{sharma2023combining,
  title={Combining Cubic Dynamical Solvers with Make/Break Heuristics to Solve SAT},
  author={Sharma, Anshujit and Burns, Matthew and Huang, Michael C},
  booktitle={26th International Conference on Theory and Applications of Satisfiability Testing (SAT 2023)},
  pages={25--1},
  year={2023},
  organization={Schloss Dagstuhl--Leibniz-Zentrum f{\"u}r Informatik}
}

@article{bohm_order--magnitude_2021,
	title = {Order-of-magnitude differences in computational performance of analog {Ising} machines induced by the choice of nonlinearity},
	volume = {4},
	copyright = {2021 The Author(s)},
	issn = {2399-3650},
	url = {https://www.nature.com/articles/s42005-021-00655-8},
	doi = {10.1038/s42005-021-00655-8},
	number = {1},
	urldate = {2023-12-21},
	journal = {Communications Physics},
	author = {Böhm, Fabian and Vaerenbergh, Thomas Van and Verschaffelt, Guy and Van der Sande, Guy},
	month = jul,
	year = {2021},
	note = {Number: 1
Publisher: Nature Publishing Group},
	pages = {1--11},
}

@inproceedings{ayodele_penalty_2022,
	address = {Cham},
	series = {Lecture {Notes} in {Computer} {Science}},
	title = {Penalty {Weights} in {QUBO} {Formulations}: {Permutation} {Problems}},
	isbn = {978-3-031-04148-8},
	shorttitle = {Penalty {Weights} in {QUBO} {Formulations}},
	doi = {10.1007/978-3-031-04148-8_11},
	booktitle = {Evolutionary {Computation} in {Combinatorial} {Optimization}},
	publisher = {Springer International Publishing},
	author = {Ayodele, Mayowa},
	editor = {Pérez Cáceres, Leslie and Verel, Sébastien},
	year = {2022},
	pages = {159--174},
}

@article{ding_langevin_2021,
	title = {Langevin {Monte} {Carlo}: random coordinate descent and variance reduction},
	volume = {22},
	issn = {1532-4435},
	shorttitle = {Langevin {Monte} {Carlo}},
	number = {1},
	journal = {The Journal of Machine Learning Research},
	author = {Ding, Zhiyan and Li, Qin},
	month = jan,
	year = {2021},
	pages = {205:9312--205:9362},
}

@article{tatsumura_scaling_2021,
	title = {Scaling out {Ising} machines using a multi-chip architecture for simulated bifurcation},
	volume = {4},
	issn = {2520-1131},
	url = {https://www.nature.com/articles/s41928-021-00546-4},
	doi = {10.1038/s41928-021-00546-4},
	number = {3},
	urldate = {2024-10-03},
	journal = {Nature Electronics},
	author = {Tatsumura, Kosuke and Yamasaki, Masaya and Goto, Hayato},
	month = mar,
	year = {2021},
	pages = {208--217},
}

@article{du_new_2025,
	title = {New advances for quantum-inspired optimization},
	volume = {32},
	issn = {1475-3995},
	url = {https://onlinelibrary.wiley.com/doi/abs/10.1111/itor.13420},
	doi = {10.1111/itor.13420},
	number = {1},
	urldate = {2024-10-06},
	journal = {International Transactions in Operational Research},
	author = {Du, Yu and Wang, Haibo and Hennig, Rick and Hulandageri, Amit and Kochenberger, Gary and Glover, Fred},
	year = {2025},
	note = {\_eprint: https://onlinelibrary.wiley.com/doi/pdf/10.1111/itor.13420},
	pages = {6--17},
}

@article{liang_information_2016,
	title = {Information flow and causality as rigorous notions \textit{ab initio}},
	volume = {94},
	copyright = {http://creativecommons.org/licenses/by/3.0/},
	issn = {2470-0045, 2470-0053},
	url = {https://link.aps.org/doi/10.1103/PhysRevE.94.052201},
	doi = {10.1103/PhysRevE.94.052201},
	number = {5},
	urldate = {2024-10-11},
	journal = {Physical Review E},
	author = {Liang, X. San},
	month = nov,
	year = {2016},
	pages = {052201},
}

@inproceedings{burns_provable_2024,
  title = {Provable {{Convergence Bounds}} for {{Hybrid Dynamical Sampling}} and {{Optimization}}},
  booktitle = {The {{Thirteenth International Conference}} on {{Learning Representations}}},
  author = {Burns, Matthew X. and Hou, Qingyuan and Huang, Michael},
  year = {2024},
  month = oct,
  urldate = {2025-07-16},
  langid = {english}
}

@article{dechant_minimum_2022,
	title = {Minimum entropy production, detailed balance and {Wasserstein} distance for continuous-time {Markov} processes},
	volume = {55},
	issn = {1751-8121},
	url = {https://dx.doi.org/10.1088/1751-8121/ac4ac0},
	doi = {10.1088/1751-8121/ac4ac0},
	number = {9},
	urldate = {2024-10-27},
	journal = {Journal of Physics A: Mathematical and Theoretical},
	author = {Dechant, Andreas},
	month = feb,
	year = {2022},
	note = {Publisher: IOP Publishing},
	pages = {094001},
}

@article{shiraishi_speed_2018,
	title = {Speed {Limit} for {Classical} {Stochastic} {Processes}},
	volume = {121},
	url = {https://link.aps.org/doi/10.1103/PhysRevLett.121.070601},
	doi = {10.1103/PhysRevLett.121.070601},
	number = {7},
	urldate = {2024-10-30},
	journal = {Physical Review Letters},
	author = {Shiraishi, Naoto and Funo, Ken and Saito, Keiji},
	month = aug,
	year = {2018},
	note = {Publisher: American Physical Society},
	pages = {070601},
}

@article{sharma_augmenting_2023,
	title = {Augmenting an electronic {Ising} machine to effectively solve boolean satisfiability},
	volume = {13},
	copyright = {2023 The Author(s)},
	issn = {2045-2322},
	url = {https://www.nature.com/articles/s41598-023-49966-6},
	doi = {10.1038/s41598-023-49966-6},
	number = {1},
	urldate = {2024-11-03},
	journal = {Scientific Reports},
	author = {Sharma, Anshujit and Burns, Matthew and Hahn, Andrew and Huang, Michael},
	month = dec,
	year = {2023},
	note = {Publisher: Nature Publishing Group},
	pages = {22858},
}

@article{nakazato_geometrical_2021,
	title = {Geometrical aspects of entropy production in stochastic thermodynamics based on {Wasserstein} distance},
	volume = {3},
	url = {https://link.aps.org/doi/10.1103/PhysRevResearch.3.043093},
	doi = {10.1103/PhysRevResearch.3.043093},
	number = {4},
	urldate = {2024-11-04},
	journal = {Physical Review Research},
	author = {Nakazato, Muka and Ito, Sosuke},
	month = nov,
	year = {2021},
	note = {Publisher: American Physical Society},
	pages = {043093},
}

@article{shiraishi_wasserstein_2024,
	title = {Wasserstein distance in speed limit inequalities for {Markov} jump processes},
	volume = {2024},
	issn = {1742-5468},
	url = {https://dx.doi.org/10.1088/1742-5468/ad5438},
	doi = {10.1088/1742-5468/ad5438},
	number = {7},
	urldate = {2024-11-04},
	journal = {Journal of Statistical Mechanics: Theory and Experiment},
	author = {Shiraishi, Naoto},
	month = jul,
	year = {2024},
	note = {Publisher: IOP Publishing},
	pages = {074003},
}

@misc{dechant_thermodynamic_2019,
	title = {Thermodynamic interpretation of {Wasserstein} distance},
	url = {http://arxiv.org/abs/1912.08405},
	doi = {10.48550/arXiv.1912.08405},
	urldate = {2024-11-05},
	publisher = {arXiv},
	author = {Dechant, Andreas and Sakurai, Yohei},
	month = dec,
	year = {2019},
	note = {arXiv:1912.08405},
}

@misc{santos_enhancing_2024,
	title = {Enhancing {Quantum} {Annealing} via entanglement distribution},
	url = {http://arxiv.org/abs/2212.02465},
	urldate = {2024-11-12},
	publisher = {arXiv},
	author = {Santos, Raúl and Buffoni, Lorenzo and Omar, Yasser},
	month = mar,
	year = {2024},
	note = {arXiv:2212.02465},
}

@inproceedings{chewi_analysis_2022,
	title = {Analysis of {Langevin} {Monte} {Carlo} from {Poincare} to {Log}-{Sobolev}},
	url = {https://proceedings.mlr.press/v178/chewi22a.html},
	urldate = {2024-11-13},
	booktitle = {Proceedings of {Thirty} {Fifth} {Conference} on {Learning} {Theory}},
	publisher = {PMLR},
	author = {Chewi, Sinho and Erdogdu, Murat A. and Li, Mufan and Shen, Ruoqi and Zhang, Shunshi},
	month = jun,
	year = {2022},
	note = {ISSN: 2640-3498},
	pages = {1--2},
}

@article{dalalyan_user-friendly_2019,
	title = {User-friendly guarantees for the {Langevin} {Monte} {Carlo} with inaccurate gradient},
	volume = {129},
	issn = {0304-4149},
	url = {https://www.sciencedirect.com/science/article/pii/S0304414918304824},
	doi = {10.1016/j.spa.2019.02.016},
	number = {12},
	urldate = {2024-11-13},
	journal = {Stochastic Processes and their Applications},
	author = {Dalalyan, Arnak S. and Karagulyan, Avetik},
	month = dec,
	year = {2019},
	pages = {5278--5311},
}

@article{panchenko_sherrington-kirkpatrick_2012,
	title = {The {Sherrington}-{Kirkpatrick} {Model}: {An} {Overview}},
	volume = {149},
	issn = {1572-9613},
	shorttitle = {The {Sherrington}-{Kirkpatrick} {Model}},
	url = {https://doi.org/10.1007/s10955-012-0586-7},
	doi = {10.1007/s10955-012-0586-7},
	number = {2},
	urldate = {2024-11-14},
	journal = {Journal of Statistical Physics},
	author = {Panchenko, Dmitry},
	month = oct,
	year = {2012},
	pages = {362--383},
}

@article{erdos_evolution_1960,
	title = {On the evolution of random graphs},
	volume = {5},
	journal = {Publ. Math. Inst. Hungary. Acad. Sci.},
	author = {Erdos, Paul and Renyi, Alfred},
	year = {1960},
	pages = {17--61},
}

@book{liptser_statistics_2001,
	address = {Berlin, Heidelberg},
	title = {Statistics of {Random} {Processes}},
	copyright = {http://www.springer.com/tdm},
	isbn = {978-3-642-08366-2 978-3-662-13043-8},
	url = {http://link.springer.com/10.1007/978-3-662-13043-8},
	urldate = {2024-11-23},
	publisher = {Springer},
	author = {Liptser, Robert S. and Shiryaev, Albert N.},
	year = {2001},
	doi = {10.1007/978-3-662-13043-8},
}

@inproceedings{jain_low-rank_2013,
	address = {New York, NY, USA},
	series = {{STOC} '13},
	title = {Low-rank matrix completion using alternating minimization},
	isbn = {978-1-4503-2029-0},
	url = {https://dl.acm.org/doi/10.1145/2488608.2488693},
	doi = {10.1145/2488608.2488693},
	urldate = {2024-12-20},
	booktitle = {Proceedings of the forty-fifth annual {ACM} symposium on {Theory} of {Computing}},
	publisher = {Association for Computing Machinery},
	author = {Jain, Prateek and Netrapalli, Praneeth and Sanghavi, Sujay},
	month = jun,
	year = {2013},
	pages = {665--674},
}

@inproceedings{yi_alternating_2014,
	title = {Alternating {Minimization} for {Mixed} {Linear} {Regression}},
	url = {https://proceedings.mlr.press/v32/yia14.html},
	urldate = {2024-12-20},
	booktitle = {Proceedings of the 31st {International} {Conference} on {Machine} {Learning}},
	publisher = {PMLR},
	author = {Yi, Xinyang and Caramanis, Constantine and Sanghavi, Sujay},
	month = jun,
	year = {2014},
	note = {ISSN: 1938-7228},
	pages = {613--621},
}

@inproceedings{kalehbasti_ising-based_2021,
	address = {Cham},
	title = {Ising-{Based} {Louvain} {Method}: {Clustering} {Large} {Graphs} with {Specialized} {Hardware}},
	isbn = {978-3-030-74251-5},
	shorttitle = {Ising-{Based} {Louvain} {Method}},
	doi = {10.1007/978-3-030-74251-5_28},
	booktitle = {Advances in {Intelligent} {Data} {Analysis} {XIX}},
	publisher = {Springer International Publishing},
	author = {Kalehbasti, Pouya Rezazadeh and Ushijima-Mwesigwa, Hayato and Mandal, Avradip and Ghosh, Indradeep},
	editor = {Abreu, Pedro Henriques and Rodrigues, Pedro Pereira and Fernández, Alberto and Gama, João},
	year = {2021},
	pages = {350--361},
}

\end{document}